\documentclass[twocolumn]{autart}    

\usepackage{graphicx}
\usepackage{epsfig}
\usepackage{amsmath}

\usepackage{amssymb}
\usepackage[noadjust]{cite}
\usepackage{bm}
\usepackage{subfigure}
\usepackage{acronym}
\usepackage{paralist}
\usepackage{float}
\usepackage{epstopdf}
\usepackage{algorithm}
\usepackage{algpseudocode}
\usepackage{mathtools}
\usepackage{multicol} 
\usepackage{accents}
\newcommand{\subparagraph}{}
\usepackage{titlesec}

\newtheorem{define}{Definition}
\newtheorem{theorem}{Theorem}
\newtheorem{lemma}{Lemma}

\newtheorem{remark}{Remark}
\newtheorem{assume}{Assumption}

\newcommand{\Bc}{\mathcal{B}}
\newcommand{\Rc}{\mathcal R}
\newcommand{\R}{\mathbb R}

\newcommand{\N}{\mathcal{N}}
\newcommand{\D}{\mathcal{D}}
\newcommand{\V}{\mathcal{V}}

\newcommand{\G}{\mathcal{G}}
\newcommand{\E}{\mathcal{E}}

\newcommand{\Xr}[2]{\mathcal{X}^{#1}_{S_{#2}}}

\newcommand{\Z}{\mathbb{Z}}
\newcommand{\Nat}{\mathbb{N}}
\newcommand{\Pc}{\mathcal{P}}
\newcommand{\Tc}{\mathcal{T}}

\newcommand{\bmx}[1]{\begin{bmatrix}#1\end{bmatrix}} 
\newcommand{\pth}[1]{\left(#1\right)} 
\newcommand{\nrm}[1]{\left \lVert#1\right \rVert} 

\newcommand{\bmxs}[1]{\begin{bsmallmatrix}#1\end{bsmallmatrix}}
\newcommand{\pmxs}[1]{\begin{psmallmatrix}#1\end{psmallmatrix}}

\newcommand{\gleq}{\preceq} 

\DeclarePairedDelimiter{\ceil}{\lceil}{\rceil}
\DeclarePairedDelimiter{\floor}{\lfloor}{\rfloor}
\DeclarePairedDelimiter{\abs}{\lvert}{\rvert}

\newcommand{\rarr}{\rightarrow} 
\newcommand{\larr}{\leftarrow} 

\makeatletter
\let\oldceil\ceil
\def\ceil{\@ifstar{\oldceil}{\oldceil*}}
\makeatother
\makeatletter
\let\oldfloor\floor
\def\floor{\@ifstar{\oldfloor}{\oldfloor*}}
\makeatother
\makeatletter
\let\oldnorm\norm
\def\norm{\@ifstar{\oldnorm}{\oldnorm*}}
\makeatother
\makeatletter
\let\oldabs\abs
\def\abs{\@ifstar{\oldabs}{\oldabs*}}
\makeatother

\theoremstyle{plain}
\newtheorem{problem}{Problem}
\newtheorem{property}{Property}

\makeatletter
\renewcommand{\ALG@beginalgorithmic}{\footnotesize}
\makeatother

\titlespacing\section{0pt}{11pt plus 4pt minus 2pt}{2pt plus 2pt minus 2pt}

\newcommand{\rmax}{{r_{\max}}}
\newcommand{\smax}{{s_{\max}}}
\newcommand{\smin}{\bar{s}_{\min}}
\newcommand{\Thr}{\Theta_r}
\newcommand{\bThr}{\bar{\Theta}_r}

\newcommand{\condA}{|\Xr{r}{1}| = |S_1|}
\newcommand{\condB}{|\Xr{r}{2}| = |S_2|}
\newcommand{\condC}[1]{|\Xr{r}{1}| + |\Xr{r}{2}| \geq #1}

\newcommand{\bsig}{\bm \sigma}
\newcommand{\delin}{\delta^{\text{in}}}
\newcommand{\maxset}{\max \pth{\Rc(S_1),\Rc(S_2)}}
\newcommand{\maxstar}{\max \pth{\Rc(S_1^*),\Rc(S_2^*)}}
\newcommand{\Pce}{\Pc_{\emptyset,\V}}

\newcommand{\simsize}{1.175}

\begin{document}

\begin{frontmatter}

\title{Determining $r$- and $(r,s)$-Robustness of Digraphs Using Mixed Integer Linear Programming \thanksref{footnoteinfo}} 

\thanks[footnoteinfo]{The authors would like to acknowledge the support of the Automotive Research Center (ARC) in accordance with Cooperative Agreement W56HZV-14-2-0001 U.S. Army TARDEC in Warren, MI, and the Award No W911NF-17-1-0526.}

\author[Umich]{James Usevitch}\ead{usevitch@umich.edu},    
\author[Umich]{Dimitra Panagou}\ead{dpanagou@umich.edu},               

\address[Umich]{Department of Aerospace Engineering, University of Michigan}  

\begin{keyword}                           
Networks, Robustness, Graph Theory, Optimization problems, Integer programming							            
\end{keyword}                             

\begin{abstract}                          
There has been an increase in the use of resilient control algorithms based on the graph theoretic properties of $r$- and $(r,s)$-robustness. These algorithms guarantee consensus of normally behaving agents in the presence of a bounded number of arbitrarily misbehaving agents if the values of the integers $r$ and $s$ are sufficiently large. However, determining an arbitrary graph's robustness is a highly nontrivial problem. This paper introduces a novel method for determining the $r$- and $(r,s)$-robustness of digraphs using mixed integer linear programming; to the best of the authors' knowledge it is the first time that mixed integer programming methods have been applied to the robustness determination problem. The approach only requires knowledge of the graph Laplacian matrix, and can be formulated with binary integer variables. Mixed integer programming algorithms such as branch-and-bound are used to iteratively tighten the lower and upper bounds on $r$ and $s$. Simulations are presented which compare the performance of this approach to prior robustness determination algorithms.
\end{abstract}

\end{frontmatter}

\section{Introduction}
\label{intro}

\vspace{-.5em}

Consensus on shared information is fundamental to the operation of multi-agent systems. In context of mobile agents, it enables formation control, rendezvous, distributed estimation, and many more objectives. Although a vast number of algorithms for consensus exist, many are unable to tolerate the presence of misbehaving agents subject to attacks or faults.
Recent years have seen an increase of attention on \emph{resilient} algorithms that are able to operate despite such misbehavior.
Many of these algorithms have been inspired by \cite{Lamport1982}, which is one of the seminal papers on consensus in the presence of adversaries; \cite{leblanc2013resilient,zhang2012robust,LeBlanc_2013_Res_Continuous}, which outline discrete- and continuous-time algorithms along with necessary and sufficient conditions for scalar consensus in the presence of Byzantine adversaries; and \cite{Vaidya2012iterative,Vaidya2013byzantine,Tseng2013iterative,Tseng2014asynchronous}, which outline algorithms for multi-agent vector consensus of asynchronous systems in the presence of Byzantine adversaries. Some of the most recent resilience-based results that draw upon these papers include state estimation \cite{Mitra2018secure,mitra2018distributedinterm,mitra2018impact,mitra2016secure}, rendezvous of mobile agents \cite{Park2017fault,park2016efficient},  output synchronization \cite{leblanc2017resilient}, simultaneous arrival of interceptors \cite{li2018robust}, distributed optimization \cite{sundaram2018distributed,su2016fault}, reliable broadcast \cite{tseng2015broadcast,zhang2012robust}, clock synchronization \cite{kikuya2017fault}, randomized quantized consensus \cite{dibaji2018resilient}, self-triggered coordination \cite{senejohnny2018resilience}, and multi-hop communication \cite{su2017reaching}.

A large number of results on network resilience are based upon the graph theoretical properties known as $r$-robustness and $(r,s)$-robustness \cite{leblanc2013resilient,zhang2012robust}. $r$-robustness and $(r,s)$-robustness
are key notions included in the sufficient conditions for convergence of several resilient consensus algorithms including the ARC-P \cite{LeBlanc_2013_Res_Continuous}, W-MSR \cite{leblanc2013resilient}, SW-MSR \cite{saldana2017resilient}, and DP-MSR \cite{dibaji2017resilient} algorithms. Given an upper bound on the global or local number of adversaries in the network, the aforementioned resilient algorithms guarantee convergence of normally behaving agents' states to a value within the convex hull of initial states if the integers $r$ and $s$ are sufficiently large.

A key challenge in implementing these resilient algorithms is that determining the $r$- and $(r,s)$-robustness of arbitrary digraphs is an NP-hard problem in general \cite{leblanc2013algorithms,zhang2015notion}.
The first algorithmic analysis of determining the values of $r$ and $s$ for arbitrary digraphs was given in \cite{leblanc2013algorithms}. The algorithms proposed in \cite{leblanc2013algorithms} employ an exhaustive search to determine the maximum values of $r$ and $s$ for a given digraph,
and have exponential complexity w.r.t. the number of nodes in the network.
In \cite{zhang2015notion} it was shown that the decision problem of determining if an arbitrary graph is $r$-robust for a given integer $r$ is coNP-complete in general.
Subsequent work has focused on methods to circumvent this difficulty, including
graph construction methods which increase the graph size while preserving given values of $r$ and $s$ \cite{leblanc2013resilient,Guerrero2016formations};
lower bounding $r$ with the isoperimetric constant and algebraic connectivity of undirected graphs \cite{shahrivar2015robustness}; and
demonstrating the behavior of $r$ as a function of certain graph properties
\cite{zhang2015notion,shahrivar2017spectral,zhao2017connectivity,
usevitch2017robust,guerrero2018design,saldana2019resilient}.
In particular, it has been shown that the $r$-robustness of some specific classes of graphs can be exactly determined in polynomial time from certain graph parameters. Examples include $k$-circulant graphs \cite{usevitch2017robust} and lattice-based formations \cite{guerrero2018design,saldana2019resilient}.
Another recent approach has used
machine learning to correlate characteristics of
certain graphs to the values of $r$ and $s$ \cite{wang2018using}, but these correlations are inherently stochastic in nature and do not provide explicit guarantees. 
Despite the impressive results of prior literature, methods to either approximate or determine exactly the $r$- and $(r,s)$-robustness of arbitrary \emph{digraphs} remain relatively rare. Methods to determine the \emph{exact} $r$- or $(r,s)$-robustness of arbitrary undirected graphs are also uncommon.
Finding more efficient or practical ways of determining the robustness of arbitrary graphs in general, and digraphs in particular, remains an open problem.

In response to this open problem, this paper introduces novel methods for determining the $r$- and $(r,s)$-robustness of digraphs and undirected graphs using mixed integer linear programming (MILP). These methods only require knowledge of the graph Laplacian matrix and are zero-one MILPs, i.e. with all integer variables being binary.
To the best of our knowledge, this is the first time the robustness determination problem has been formulated in this way. These results connect the problem of graph robustness determination to the extensive and well-established literature on integer programming and  linear programming.

This paper makes the following specific contributions:
\begin{enumerate}
    \item We present a method to determine the maximum integer for which a nonempty, nontrivial, simple digraph is $r$-robust using mixed integer linear programming.
    \item We present a method which determines the $(r,s)$-robustness of a digraph using  linear programming. Here, the $(r,s)$-robustness of a digraph refers to the maximal $(r,s)$ integer pair according to a lexicographical order for which a given digraph is $(r,s)$-robust, as first described in \cite{leblanc2013algorithms}. Furthermore, we show that our method can also determine the maximum integer $F$ for which a digraph is $(F+1,F+1)$-robust, which is not considered in \cite{leblanc2013algorithms}.
     \item We present two mixed integer linear programs whose optimal values provide lower and upper bounds on the maximum $r$ for which a nonempty, nontrivial, simple digraph is $r$-robust. These two formulations exhibit a lower complexity than the method in the first contribution described above.
\end{enumerate}
The contributions of this paper provide several advantages. First, expressing the robustness determination problem in MILP form allows for approximate \emph{lower} bounds on a given digraph's $r$-robustness to be iteratively tightened using algorithms such as branch-and-bound. Lower bounds on the maximum value of $s$ for which a given digraph is $(r,s)$ robust (for a given nonnegative integer $r$) can also be iteratively tightened using the approach in this paper. 
Prior algorithms are only able to tighten the upper bound on the maximum robustness for a given digraph or undirected graph.
Second, this formulation enables commercially available solvers such as Gurobi or MATLAB's \emph{intlinprog} to be used to find the maximum robustness of any digraph. Finally, experimental results using this new formulation suggest a reduction in computation time as compared to the centralized algorithm proposed in \cite{leblanc2013algorithms}.

Part of this paper was previously submitted as a conference paper \cite{usevitch2019determining}. The extensions to the conference version include the following:
\begin{itemize}
    \item The more general case of determining $(r,s)$-robustness is considered.
    \item Two optimization problems with reduced-dimension binary vector variables are given whose optimal values provide a lower bound and an upper bound, respectively, on the maximum value of $r$ for which a graph is $r$-robust.
\end{itemize}

This paper is organized as follows: notation is presented in Section \ref{sec:notation},
the problem formulation is given in Section \ref{sec:problemformulation}, determining the $r$-robustness of digraphs is treated in Section \ref{sec:rrobust}, determining the values of $s$ for which a digraph is $(r,s)$-robust for a given $r$ is treated in Section \ref{sec:rsrobust}, methods to obtain upper and lower bounds on the maximum $r$ for which a digraph is $r$-robust are presented in Section \ref{sec:reducecomplex}, 
a brief discussion about the advantages of the MILP formulations is given in Section \ref{sec:discussion},
simulations demonstrating our algorithms are presented in Section \ref{sec:sim}, and a brief conclusion is given in \ref{sec:conclusion}.

\vspace{-1em}

\section{Notation}
\label{sec:notation}

The sets of real numbers and integers are denoted $\R$ and $\Z$, respectively. The sets of nonnegative real numbers and integers are denoted $\R_+$ and $\Z_+$, respectively. $\R^n$ denotes an $n$-dimensional vector space over the field $\R$, $\Z^n$ represents the set of $n$ dimensional vectors with integer entries, and $\{0,1\}^n$ represents a binary vector of dimension $n$. Scalars are denoted in normal text (e.g. $x \in \R$) while vectors are denoted in bold (e.g. $\bm x \in \R^n$).
The notation $x_i$ denotes the $i$th entry of vector $\bm x$.

The inequality symbol $\gleq$ denotes a componentwise inequality between vectors; i.e. for $\bm x,\bm y \in \R^n$, $\bm x \gleq \bm y$ implies $x_i \leq y_i\ \forall i \in \{1,\ldots,n\}$. A vector of ones is denoted $\bm 1$, and a vector of zeros is denoted $\bm 0$, where the length of each vector will be implied by the context.
The union, intersection, and set complement operations are denoted by $\cup,\ \cap$, and $\setminus$, respectively.
The cardinality of a set $S$ is denoted as $|S|$, and the empty set is denoted $\emptyset$. The infinity norm of a vector is denoted $\nrm{\cdot}_\infty$. The notations $C(n,k) = \pmxs{n \\ k} = n!/(k!(n-k)!)$ are both used in this paper to denote the binomial coefficient with $n,k \in \Z_+$. Given a set $S$, the power set of $S$ is denoted $\Pc(S) = \{A : A \subseteq S \}$. Given a function $f : D \rarr R$, the image of a set $A \subseteq D$ under $f$ is denoted $f(A)$. Similarly, the preimage of $B \subseteq R$ under $f$ is denoted $f^{-1}(B)$. The logical OR operator, AND operator, and NOT operator are denoted by $\lor, \land, \lnot$, respectively. The lexicographic cone is defined as $K_{lex} = \{0\} \cup \{\bm x \in \R^n : x_1 = \ldots = x_{k} = 0,\ x_{k+1} > 0 \}$ for some $0 \leq k < n$. The lexicographic ordering on $\R^n$ is defined as $\bm x \leq_{lex} \bm y$ if and only if $\bm y - \bm x \in K_{lex}$, with $\bm x, \bm y \in \R^n$ \cite[Ch. 2]{boyd2004convex}.

A directed graph (digraph) is denoted as $\D = (\V,\E)$, where $\V = \{1,\ldots,n\}$ is the set of indexed vertices and $\E$ is the edge set. This paper will use the terms vertices, agents, and nodes interchangeably.
A directed edge is denoted $(i,j)$, with $i,j \in \V$, meaning that agent $j$ can receive information from agent $i$. The set of in-neighbors for an agent $j$ is denoted $\N_j = \{i \in \V : (i,j) \in \E \}$. The minimum in-degree of a digraph $\D$ is denoted $\delta^{in}(\D) = \min_{j \in \V} |\N_j|$.
Occasionally, $\G = (\V,\E)$ will be used to denote an undirected graph, i.e. a digraph in which $(i,j) \in \E \iff (j,i) \in \E$.
The graph Laplacian $L$ for a digraph (or undirected graph) is defined as follows, with $L_{i,j}$ denoting the entry in the $i$th row and $j$th column:
\begin{equation}
\label{eq:Laplacian}
    L_{i,j} = \begin{cases}
        |\N_i| & \text{if } i = j, \\
        -1 & \text{if } j \in \N_i, \\
        0 & \text{if } j \notin \N_i.
    \end{cases}
\end{equation}

\vspace{-1em}

\section{Problem Formulation}
\label{sec:problemformulation}

The notions of $r$- and $(r,s)$-robustness are graph theoretical properties used to describe the communication topologies of multi-agent networks. Examples of such networks include stations in a power grid, satellites in formation, or a group of mobile robots. In these networks, edges model the ability for one agent $i$ to transmit information to another agent $j$.
Prior literature commonly considers \emph{simple} digraphs, which have no repeated edges or self edges \cite{leblanc2012consensus, leblanc2013resilient, leblanc2012low, Vaidya2012iterative,vaidya2014iterative}. More specifically, simple digraphs satisfy $(i,i) \notin \E\ \forall i \in \V$, and if the directed edge $(i,j) \in \E$, then it is the only directed edge from $i$ to $j$.
Prior work also commonly considers nonempty and nontrivial graphs, where $|V| > 1$. 
\begin{assume}
\label{assume:simple}
This paper considers nonempty, nontrivial, simple digraphs.
\end{assume}

The property of $r$-robustness is based upon the notion of $r$-reachability. The definitions of $r$-reachability and $r$-robustness are as follows:

\begin{define}[\cite{leblanc2013resilient}]
\label{def:rreach}
Let $r \in \Z_+$ and $\D=(\V,\E)$ be a digraph. A nonempty subset $S \subset \V$ is $r$-reachable if $\exists i \in S$ such that $|\N_i \backslash S| \geq r$.
\end{define}
\begin{define}[\cite{leblanc2013resilient}]
\label{def:rrobust}
Let $r \in \Z_+$. A nonempty, nontrivial digraph $\D = (\V,\E)$ on $n$ nodes $(n \geq 2)$ is $r$-robust if for every pair of nonempty, disjoint subsets of $\V$, at least one of the subsets is $r$-reachable. By convention, the empty  graph $(n = 0)$ is 0-robust and the trivial graph $(n=1)$ is 1-robust.
\end{define}

The property of $(r,s)$-robustness is based upon the notion of $(r,s)$-reachability. The definitions of $(r,s)$-reachability and $(r,s)$-robustness are as follows:

\begin{define}[\cite{leblanc2013resilient}]
Let $\D = (\V,\E)$ be a nonempty, nontrivial, simple digraph on $n \geq 2$ nodes. Let $r,s \in \Z_+$, $0 \leq s \leq n$. Let $S$ be a nonempty subset of $\V$, and define the set $\Xr{r}{} = \{j \in S : |\N_j \backslash S| \geq r \}$. We say that $S$ is an $(r,s)$-reachable set if there exist $s$ nodes in $S$, each of which has at least $r$ in-neighbors outside of $S$. More explicitly, $S$ is $(r,s)$-reachable if $|\Xr{r}{}| \geq s$.
\end{define}

\begin{define}[\cite{leblanc2013resilient}]
\label{def:rsrobust}
Let $r,s \in \Z_+$, $0 \leq s \leq n$. Let $\D = (\V,\E)$ be a nonempty, nontrivial, simple digraph on $n \geq 2$ nodes. Define $\Xr{r}{} = \{j \in S : |\N_j \backslash S| \geq r \}$, $S \subset \V$.
The digraph $\D$ is $(r,s)$-robust if for every pair of nonempty, disjoint subsets $S_1, S_2 \subset \V$, at least one of the following conditions holds:
\begin{enumerate}[A)]
    \item $|\Xr{r}{1}| = |S_1|$,
    \item $|\Xr{r}{2}| = |S_2|$,
    \item $|\Xr{r}{1}| + |\Xr{r}{2}| \geq s$.
\end{enumerate}
\end{define}
The properties of $r$- and $(r,s)$-robustness are used to quantify the ability of several resilient consensus algorithms to guarantee convergence of normally behaving agents in the presence of Byzantine and malicious adversaries, collectively referred to in this paper as \emph{misbehaving} agents \cite{leblanc2013resilient,leblanc2017resilient,dibaji2017resilient,
saldana2017resilient,LeBlanc_2013_Res_Continuous}. Larger values of $r$ and $s$ generally imply the ability of networks applying these resilient algorithms to tolerate a greater number of misbehaving agents in the network.
For a more detailed explanation of the properties of $r$-robustness and $(r,s)$-robustness, the reader is referred to \cite{leblanc2012consensus,leblanc2013resilient,zhang2015notion}.

\begin{figure}
	\centering
	\includegraphics[width = .8\columnwidth]{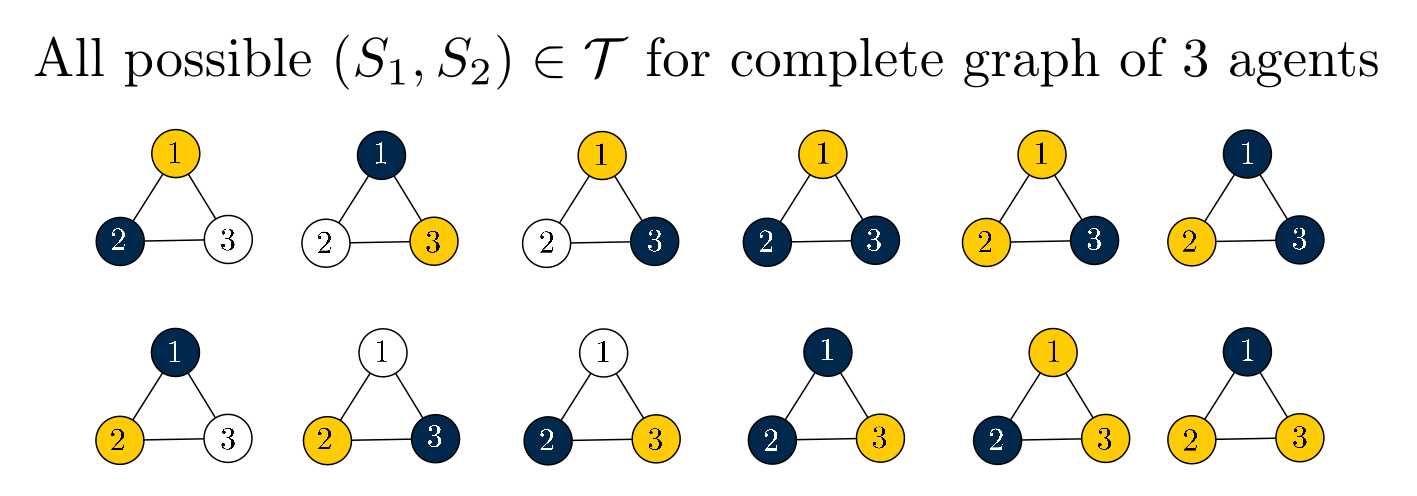}
	\caption{
	{\scriptsize	
	Depiction of all 12 possible $(S_1,S_2)$ elements in $\Tc$ for a complete graph $\D$ of 3 agents. Each graph represents a different possible way of dividing $\D$ into sets $S_1$ and $S_2$. In each individual graph, yellow agents are in $S_1$, blue agents are in $S_2$, and white agents are in neither $S_1$ nor $S_2$. 
	}
	}
	\label{fig:s1s2example}
\end{figure}

It should be clear from Definitions \ref{def:rrobust} and \ref{def:rsrobust} that determining $r$ and $(r,s)$-robustness for a digraph $\D = (\V,\E)$ by using an exhaustive search method is a combinatorial problem, which involves checking the reachabilities of all nonempty, disjoint subsets of $\V$. For notational purposes, we will denote the set of all possible pairs of nonempty, disjoint subsets of $\V$ as $\mathcal{T} \subset \Pc(\V) \times \Pc(\V)$. More explicitly, $\Tc$ is defined as
\begin{align}
        \mathcal{T} =  \big\{ &(S_1,S_2) \in \Pc(\V) \times \Pc(\V) : |S_1| > 0,\ |S_2| > 0,\  \nonumber \\
        &|S_1 \cap S_2| = 0 \big\}. \label{eq:Tdef}
\end{align}
It was shown in \cite{leblanc2013algorithms} that $|\mathcal{T}| = \sum_{p=2}^n \pmxs{n \\ p} (2^{p} -2)$.\footnote{Since $(S_1,S_2) \in \mathcal{T} \implies (S_2,S_1) \in \mathcal{T}$, the total number of \emph{unique} nonempty, disjoint subsets is $(1/2)|\mathcal{T}|$, denoted as $R(n)$ in \cite{leblanc2013algorithms}.} As a simple example, Figure \ref{fig:s1s2example} depicts all elements of $\Tc$ for a graph of 3 agents, i.e. all possible ways to choose two nonempty, disjoint subsets from the graph.

When considering a particular digraph $\D$, there may be multiple values of $r$ for which $\D$ is $r$-robust. Similarly, there may be multiple values of $r$ and $s$ for which $\D$ is $(r,s)$-robust.
The following properties of robust graphs demonstrate this characteristic. Note that $r$-robustness is equivalent to $(r,1)$-robustness; i.e. $\D$ is $r$-robust if and only if it is $(r,1)$-robust \cite[Property 5.21]{leblanc2013algorithms} \cite[Section VII-B]{leblanc2013resilient}.
\begin{property}[\cite{leblanc2012thesis}, Prop. 5.13]
\label{prop:rsimplies}
    Let $\D$ be an arbitrary, simple digraph on $n$ nodes. Suppose $\D$ is $(r,s)$-robust with $r \in \Nat$ and $s \in \{0,\ldots,n\}$. Then $\D$ is also $(r',s')$-robust $\forall r' :\ 0 \leq r' \leq r$ and $\forall s' :\ 1 \leq s' \leq s$.
\end{property}

\begin{property}[\cite{leblanc2012thesis}, Prop. 5.20]
\label{prop:rsplus}
Let $\D$ be an arbitrary, simple digraph on $n$ nodes. Suppose $\D$ is $(r,s)$-robust with $r \in \Nat$ and $s \in \{1,\ldots,n\}$. Then $\D$ is $(r-1,s+1)$-robust.
\end{property}
As an example, if a digraph $\D_1$ is $4$-robust, it is $(4,1)$-robust, and therefore by Property \ref{prop:rsimplies} it is also simultaneously 3-robust, 2-robust, and 1-robust. In addition, if a digraph $\D_2$ is $(5,4)$-robust, then it is simultaneously $(r',s')$-robust for all integers $0 \leq r' \leq 5$ and $0 \leq s' \leq 4$. Moreover, by Property \ref{prop:rsplus}, $\D_2$ is also $(4,5)$-robust, $(3,6)$-robust, $(2,7)$-robust, and $(1,8)$-robust. For notational purposes, we denote the set of all values for which a digraph $\D$ is $(r,s)$-robust as $\Theta$, where $\Theta \subset \Z_+ \times \Z_+$.
By Definition \ref{def:rsrobust}, $\Theta$ is explicitly defined as
\begin{align}
\label{eq:Thetadef}
\Theta = \{&(r,s) \in \Z_+ \times \Z_+ : \forall(S_1,S_2) \in \Tc,
(|\Xr{r}{1}| = |S_1|) \text{ or } \nonumber\\
&(|\Xr{r}{2}| = |S_2|) \text{ or } (|\Xr{r}{1}| + |\Xr{r}{2}| \geq s) \}.
\end{align}
Note that the conditions of \eqref{eq:Thetadef} are simply an alternate way of expressing the conditions of Definition \ref{def:rsrobust}.

To characterize the resilience of graphs however, prior literature has generally been concerned with only a few particular values of $r$ and $s$ for which a given digraph is $r$- or $(r,s)$-robust. For $r$-robustness, the value of interest is the maximum integer $r$ for which the given digraph is $r$-robust.
\begin{define}
\label{rmk:rho}
	We denote the maximum integer $r$ for which a given digraph $\D$ is $r$-robust as $\rmax(\D) \in \Z_+$.
\end{define}
Several resilient algorithms guarantee convergence of the normal agents when the adversary model is $F$-total or $F$-local in scope,\footnote{An $F$-total adversary model implies that there are at most $F \in \Z_+$ misbehaving agents in the entire network. An $F$-local adversary model implies that each normal agent has at most $F$ misbehaving agents in its in-neighbor set.} and the digraph is $(2F+1)$-robust. The value of $\rmax(\D)$ therefore determines the maximum adversary model under which these algorithms can operate successfully. Furthermore, all other values of $r$ for which a digraph $\D$ is $r$-robust can be determined from $\rmax(\D)$ by using Property \ref{prop:rsimplies}.

For $(r,s)$-robustness, there are two $(r,s)$ pairs of interest. The authors of \cite{leblanc2013algorithms} order the elements of $\Theta$ using a lexicographical total order, where elements are ranked by $r$ value first and $s$ value second.
More specifically, $(r_1,s_1) \leq_{lex} (r_2,s_2)$ if and only if $\bmxs{r_2 - r_1 \\ s_2 - s_1} \in K_{lex}$, where $K_{lex}$ is the lexicographic cone defined in Section \ref{sec:notation}.
Their algorithm $DetermineRobustness$ finds the maximum element of $\Theta$ with respect to this order. 
For notational clarity, we denote this maximum element as $(r^*,s^*) \in \Theta$. 
\begin{define}
\label{def:rstarsstar}
Let $\Theta$ be defined as in \eqref{eq:Thetadef}. The element \emph{$(r^*,s^*)$} is defined as the maximum element of $\Theta$ under the lexicographical order on $\R^2$.
\end{define}
The other $(r,s)$ pair of interest is $(F_{\max} + 1, F_{\max}+1)$, where $F_{\max} = \max (\{F \in \Z_+ : (F+1,F+1) \in \Theta\})$. Several resilient algorithms guarantee convergence of the normally behaving agents when the (malicious \cite{leblanc2013resilient}) adversary model is $F$-total in scope and the digraph is $(F+1,F+1)$-robust. The value $F_{\max}$ determines the maximum malicious adversary model under which these algorithms can operate successfully. The value of $(F_{\max} + 1, F_{\max}+1)$ does not always coincide with the $(r^*,s^*)$-robustness of the digraph. A simple counterexample is given in Figure 2, where the $(r^*,s^*)$-robustness of the graph is $(2,1)$ but the value of $(F_{\max} + 1, F_{\max}+1)$ is equal to $(1,1)$.

The purpose of this paper is to present methods using mixed integer linear programming to determine $\rmax(\D)$, the $(r^*,s^*)$-robustness of $\D$, and the $(F_{\max} + 1, F_{\max}+1)$-robustness of $\D$ for any nonempty, nontrivial, simple digraph $\D$.
\begin{problem}
\label{prob:FirstProblem}
Given an arbitrary nonempty, nontrivial, simple digraph $\D$, determine the value of $\rmax(\D)$.
\end{problem}

\begin{problem}
\label{prob:SecondProblem}
Given an arbitrary nonempty, nontrivial, simple digraph $\D$, determine the $(r^*,s^*)$-robustness of $\D$.
\end{problem}

\begin{problem}
\label{prob:ThirdProblem}
Given an arbitrary nonempty, nontrivial, simple digraph $\D$, determine the $(F_{\max}+1, F_{\max}+1)$-robustness of $\D$. 
\end{problem}

Finally, the values of $r$ for which a digraph can be $r$-robust lie within the interval $0 \leq r \leq \ceil{n/2}$ \cite[Property 5.19]{leblanc2012thesis}. Since $r$-robustness is equivalent to $(r,1)$-robustness,
the values of $r$ for which a graph can be $(r,s)$-robust fall within the same interval. The values of $s$ for which a digraph can be $(r,s)$-robust lie within the interval $1 \leq s \leq n$.\footnote{Footnote 8 in \cite{leblanc2013resilient} offers an excellent explanation for restricting $s$ to this range by convention.} However, we will use an abuse of notation by denoting a graph as $(r,0)$-robust for a given $r \in \Z_+$ if the graph is not $(r,1)$-robust.

\begin{figure}
\includegraphics[width=\columnwidth]{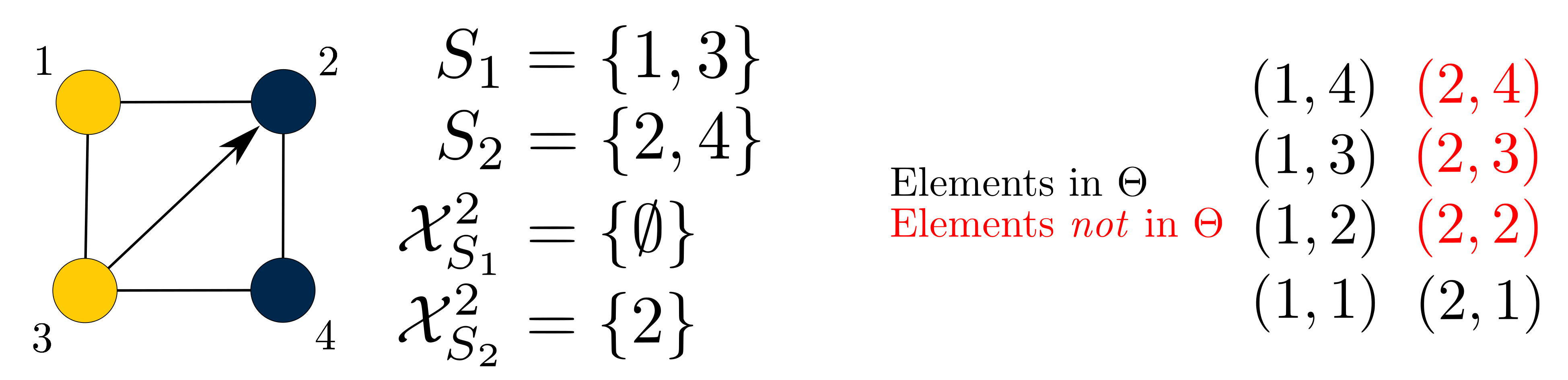}
\begin{caption}
{\scriptsize
{An} example of the elements of $\Theta$ for a digraph $\D_1$. Since $|\V| = 4$, the possible values of $r$ and $s$ for which the digraph is $(r,s)$-robust fall within the range $0 \leq r \leq 2$, $1 \leq s \leq 4$. One possible pair of subsets $S_1$ and $S_2$ is depicted, which satisfies $|\Xr{2}{1}| \neq |S_1|$, $|\Xr{2}{2}| \neq |S_2|$, $|\Xr{2}{1}| = 0$ and $|\Xr{2}{2}| = 1$. By Definition \ref{def:rsrobust}, $\D_1$ therefore cannot be $(2,2)$-robust, $(2,3)$-robust, or $(2,4)$-robust. 
}
\label{fig:rsexample}
\end{caption}
\end{figure}

\vspace{-1em}

\section{Determining $r$-Robustness using Mixed Integer Linear Programming}
\label{sec:rrobust}

In this section we will demonstrate a method for solving Problem \ref{prob:FirstProblem} using a mixed integer linear program (MILP) formulation.
An MILP will be presented whose optimal value is equal to $\rmax(\D)$ for any given nonempty, nontrivial, simple digraph $\D$. 

First, an equivalent way of expressing the maximum robustness $\rmax(\D)$ of a digraph $\D$ is derived. This equivalent expression will clarify how $\rmax(\D)$ can be determined by means of an optimization problem.
Given an arbitrary, simple digraph $\D=(\V,\E)$ and a subset $S \subset V$, the reachability function $\Rc : \Pc(\V) \rarr \Z_+$ is defined as follows:
\begin{align}
\label{eq:reachmax}
    \Rc(S) &= \begin{cases}
    \max_{i \in S } |\N_i \backslash S |, & \text{if } S \neq \{\emptyset\}, \\
    0, & \text{if } S = \{\emptyset\}.
    \end{cases}
\end{align}
In other words, the function $\Rc(S)$ returns the maximum $r$ for which $S$ is $r$-reachable.
Using this function, the following Lemma presents an optimization formulation which yields $\rmax(\D)$:

\begin{lemma}
\label{lem:rrobalt}
    Let $\D = (\V,\E)$ be an arbitrary nonempty, nontrivial, simple digraph with $|\V| = n$. Let $\rmax(\D)$ be defined as in Definition \ref{rmk:rho}. The following holds:
    \begin{align}
    \label{eq:rrobalt}
            \rmax(\D) =& \underset{S_1,S_2 \in \Pc(\V)}{\min}
            & & \max \pth{\Rc(S_1),\Rc(S_2)} \nonumber \\
            & \hspace{-2.5em}\text{\textup{subject to}}
            & & \hspace{-2.5em}|S_1| > 0,\ |S_2| > 0,\ |S_1 \cap S_2| = 0.
    \end{align}
\end{lemma}

\vspace{-3em}

\begin{pf}
	Note that $S_1,S_2 \in \Pc(\V)$ and the three constraints of the RHS of \eqref{eq:rrobalt} imply that the feasible set consists of all subsets $S_1,S_2$ such that $(S_1,S_2) \in \Tc$, as per \eqref{eq:Tdef}.
    In addition, $\maxset = m$ implies $\Rc(S_1) = m$ or $\Rc(S_2) = m$. Let $(S_1^*, S_2^*)$ be a minimizer of \eqref{eq:rrobalt}. Then $\maxstar \leq \maxset$ $\forall (S_1,S_2) \in \Tc$. Therefore $\forall (S_1,S_2) \in \Tc,$ either $\Rc(S_1) \geq \max\big(\Rc(S_1^*),$ $\Rc(S_2^*)\big)$ 
or $\Rc(S_2) \geq \maxstar$. This satisfies the definition of $r$-robustness as per Definition \ref{def:rrobust}, therefore $\D$ is at least $\maxstar$-robust. This implies $\rmax(\D) \geq \maxstar$.

    We next show that $\rmax(\D) = \maxstar$. We prove by contradiction. Recall from Definition \ref{rmk:rho} that $\rmax(\D)$ is the maximum integer $r$ for which $\D$ is $r$-robust, which means $\D$ is $\rmax(\D)$-robust by definition.
Suppose $\rmax(\D) > \maxstar$. This implies $\Rc(S_1^*) < \rmax(\D)$ and $\Rc(S_2^*) < \rmax(\D)$. Since the nonempty, disjoint subsets $(S_1^*,S_2^*) \in \Tc$ satisfy $\Rc(S_1^*) < \rmax(\D)$ and $\Rc(S_2^*) < \rmax(\D)$, by the negation of Definition \ref{def:rrobust} this implies that $\D$ is \emph{not} $\rmax(\D)$-robust.
However, this contradicts the definition of $\rmax(\D)$ being the largest integer for which $\D$ is $r$-robust (Definition \ref{rmk:rho}).
This provides the desired contradiction; therefore $\rmax(\D) = \maxstar$. \hspace*{\fill} $\blacksquare$
\end{pf}

\begin{remark}
    \label{rmk:implicit}
    Using the definition of $\mathcal{T}$ in \eqref{eq:Tdef}, the constraints on the RHS of \eqref{eq:rrobalt} can be made implicit \cite[section 4.1.3]{boyd2004convex} as follows:
    \begin{equation}
    \label{eq:rimplicit}
        \begin{aligned}
            \rmax(\D) =& \underset{(S_1,S_2) \in \mathcal{T}}{\min}
            & & \max \pth{\Rc(S_1),\Rc(S_2)}.
        \end{aligned}
    \end{equation}
\end{remark}

We demonstrate next that the objective function of \eqref{eq:rrobalt} can be expressed as a function of the network Laplacian matrix.
Recall that $n = |\V|$ and that $\{0,1\}^n$ represents a binary vector of dimension $n$. The indicator vector $\bm \sigma(\cdot) : \Pc(\V) \rarr \{0,1\}^{n}$ is defined as follows: for all $S \in \Pc(\V)$,
\begin{align}
\label{eq:sigmadef}
    \sigma_j(S) &= \begin{cases}
        1 & \text{if } j \in S \\
        0 & \text{if } j \notin S
    \end{cases},\ j = \{1,\ldots,n \}.
\end{align}
In other words the $j$th entry of $\bm \sigma(S)$ is 1 if the node with index $j$ is a member of the set $S \in \Pc(\V)$, and zero otherwise. It is straightforward to verify that $\bm \sigma : \Pc(\V) \rarr \{0,1\}^n$ is a bijection. Therefore given $\bm x \in \{0,1\}^n$, the set $\bm \sigma^{-1}(\bm x) \in \Pc(\V)$ is defined by $\bm \sigma^{-1}(\bm x) = \{j \in \V : x_j = 1 \}$. Finally, observe that for any $S \in \Pc(\V)$, $|S| = \bm 1^T \bm \sigma(S)$.
The following Lemma demonstrates that for any $S \in \Pc(\V)$, the function $\Rc(S)$ can be determined as an affine function of the network Laplacian matrix and the indicator vector of $S$:

\begin{lemma}
\label{lem:Ljsig}
    Let $\D= (\V,\E)$ be an arbitrary nonempty, nontrivial, simple digraph, let $L$ be the Laplacian matrix of $\D$, and let $S \in \Pc(\V)$. Then the following holds for all $j \in \{1,\ldots,n\}$:
    \begin{align}
    	\label{eq:Ljsig}
        L_j \bm \sigma(S) &= \begin{cases}
            |\N_j \backslash S|, & \text{if } j \in S, \\
            -|\N_j \cap S|, & \text{if } j \notin S,
        \end{cases}
    \end{align}
    where $L_j$ is the $j$th row of $L$. Furthermore,
    \begin{align}
    	\label{eq:RSLjsig}
    	\Rc(S) = \max_j L_j \bm \sigma(S),\ j \in \{1,\ldots,n\}.
    \end{align}
\end{lemma}
\vspace{-2.5em}
\begin{pf}
The term $\bm \sigma(S)$ is shortened to $\bm \sigma$ for brevity. Recall that the entry in the $j$th row and $i$th column of $L$ is denoted $L_{j,i}$. The definition of $L$ from \eqref{eq:Laplacian} implies
    \begin{align}
    	L_j \bm \sigma &= (L_{j,j}) \sigma_j + \sum_{q \in \{1,\ldots,n\} \backslash j } (L_{j,q}) \sigma_q  \nonumber \\
         &= |\N_j| \sigma_j   - \sum_{q \in \N_j \cap S } \sigma_q  - \sum_{q \in \N_j \backslash S } \sigma_q. \label{eq:Ljsummation}
    \end{align}
Since by \eqref{eq:sigmadef}, $q \in S$ implies $\sigma_q = 1$, the term $\sum_{q \in \N_j \cap S } \sigma_q = |\N_j \cap S|$. In addition, since $q \notin S$ implies $\sigma_q = 0$, the term $\sum_{q \in \N_j \backslash S } \sigma_q = 0$. By this, equation \eqref{eq:Ljsummation} simplifies to $L_j \bm  \sigma  = |\N_j| \sigma_j   - |\N_j \cap S |$.

The value of the term $|\N_j| \sigma_j$ depends on whether $j \in S$ or $j \notin S$. If $j \in S$, then $\sigma_j = 1$, implying $ L_j \bm  \sigma  = |\N_j| - |\N_j \cap S | = \pth{|\N_j \cap S | + |\N_j \backslash S |} -  |\N_j \cap S | = |\N_j \backslash S |$.
If $j \notin S $, then $\sigma_j = 0$ implying $L_j \bm  \sigma  = -|\N_j \cap S |$. This proves the result for equation \eqref{eq:Ljsig}.

To prove \eqref{eq:RSLjsig}, we first consider nonempty sets $S \in \Pc(\V) \backslash \{\emptyset\}$. By the results above and \eqref{eq:reachmax}, the maximum reachability of any $S \in \Pc(\V) \backslash \{ \emptyset \}$ is
\begin{equation}
    \Rc(S) = \max_{j \in  S} |\N_j \backslash S| = \max_{j \in S} (L_j \bm \sigma(S)). \label{eq:maxreachabilityR}
\end{equation}
By its definition, $\Rc(S) \geq 0$. Observe that
if $j \in S$ then $L_j \bm \sigma(S) = |\N_j \backslash S| \geq 0$, implying $\max_{j \in S} L_j \bm \sigma(S) \geq 0$.
Conversely,
if an agent $j$ is \emph{not} in the set $S$, then the function $L_j \bm \sigma(S)$ takes the nonpositive value $-|\N_j \cap S|$. This implies $\max_{j \notin S} L_j \bm \sigma(S) \leq 0$.
By these arguments, we therefore have $\max_{j \notin S} L_j \bm \sigma(S) \leq 0 \leq \max_{j \in S} L_j \bm \sigma(S)$, which implies
\begin{align}
	\max_{j \in \{1,\ldots,n \}} L_j \bm \sigma(S) &= \max\pth{(\max_{j \in S} L_j \bm \sigma(S)), (\max_{j \notin S} L_j \bm \sigma (S))} \nonumber \\
	&= \max_{j \in S} L_j \bm \sigma(S). \label{eq:finalmaxfunctions}
\end{align}
Therefore by equations \eqref{eq:finalmaxfunctions} and \eqref{eq:maxreachabilityR}, the maximum reachability of $S$ is found by the expression
\begin{equation}
    \label{eq:reachS}
    \Rc(S) = \max_j (L_j \bm \sigma(S)),\ j \in \{1,\ldots,n\}.
\end{equation}
Lastly, if $S = \emptyset$, then by \eqref{eq:reachmax} we have $\Rc(S) = 0$. In addition, $\bm \sigma(S) = 0$, implying that $\max_j L_j \bm \sigma(S) = 0 = \Rc(S),\ j \in \{1,\ldots,n\}$. \hspace*{\fill} $\blacksquare$
\end{pf}

Using Lemma \ref{lem:Ljsig}, it will next be shown that the objective function of \eqref{eq:rrobalt} can be rewritten as the maximum over a set of affine functions:

\begin{lemma}
\label{lem:maxequal}
Consider an arbitrary, nonempty, nontrivial, simple digraph $\D = (\V,\E)$. Let $L$ be the Laplacian matrix of $\D$, and let $L_i$ be the $i$th row of $L$. Let $\Tc$ be defined as in \eqref{eq:Tdef}. Then for all $(S_1,S_2) \in \Tc$ the following holds:
\begin{align}
    &\max \pth{\Rc(S_1),\Rc(S_2)} = \nonumber \\
     &\max \pth{ \max_{i \in \{1,\ldots,n\}} \pth{L_i \bm \sigma(S_1)}, \max_{j \in \{1,\ldots,n\}} \pth{L_j \bm \sigma(S_2)} } \nonumber
\end{align}
\end{lemma}

\vspace{-2em}

\begin{pf}
By Lemma \ref{lem:Ljsig}, $\Rc(S_1) = \max_i L_i \bm \sigma(S_1)$ and $\Rc(S_2) = \max_j L_j \bm \sigma(S_2)$ for $i,j \in \{1,\ldots,n\}$. The result follows. \hspace*{\fill} $\blacksquare$
\end{pf}
\vspace{-.5em}
From Lemma \ref{lem:rrobalt}, Lemma \ref{lem:maxequal}, and Remark \ref{rmk:implicit},
we can immediately conclude that $\rmax(\D)$ satisfies
\begin{align}
    \label{eq:rrobalt5000}
            &\rmax(\D) = \nonumber \\
            &\underset{(S_1,S_2) \in \Tc}{\min}
             \max \big(\max_i \pth{L_i \bm \sigma(S_1)}, \max_j \pth{L_j \bm \sigma(S_2)} \big).
\end{align}
Note that the terms $\bm \sigma(S_1)$ and $\bm \sigma(S_2)$ are each $n$-dimensional binary vectors. Letting $\bm b^1 = \bm \sigma(S_1)$ and $\bm b^2 = \bm \sigma(S_2)$, the objective function of \eqref{eq:rrobalt5000} can be written as $\max \big(\max_i \pth{L_i \bm b^1}, \max_j \pth{L_j \bm b^2} \big)$. Every pair $(S_1,S_2) \in \Tc$ can be mapped into a pair of binary vectors $(\bm b^1, \bm b^2)$ by the function $\Sigma : \Tc \rarr \{0,1\}^n \times \{0,1\}^n$, where $\Sigma(S_1,S_2) = (\bm \sigma(S_1), \bm \sigma(S_2)) = (\bm b^1, \bm b^2)$.
By determining the image of $\Tc$ under $\Sigma(\cdot,\cdot)$, the optimal value of \eqref{eq:rrobalt5000} can be found by minimizing over pairs of binary vectors $(\bm b_1, \bm b_2) \in \Sigma(\Tc)$ directly. Using binary vector variables instead of set variables $(S_1,S_2)$ will allow \eqref{eq:rrobalt5000} to be written directly in an MILP form.
%
%
Towards this end, the following Lemma defines the set $\Sigma(\Tc)$:

\begin{lemma}
\label{lem:bijec}
    Let $\D = (\V,\E)$ be an arbitrary nonempty, nontrivial, simple digraph, and let $\Tc$ be defined as in \eqref{eq:Tdef}.
    Define the function $\Sigma: \mathcal{T} \rarr \{0,1\}^n \times \{0,1\}^n$ as
    \begin{equation}
    \label{eq:Sigma}
        \Sigma(S_1,S_2) = (\bm \sigma(S_1),\bm \sigma(S_2)),\ (S_1,S_2) \in \Tc.
    \end{equation}
    Define the set $\Bc \subset \{0,1\}^n \times \{0,1\}^n$ as
    \begin{align}
        \mathcal{B} = \bigg\{&  (\bm b^1, \bm b^2) \in \{0,1 \}^n \times \{0,1 \}^n : 1 \leq \bm 1^T \bm b^1 \leq (n-1), \nonumber \\
         &1 \leq \bm 1^T \bm b^2 \leq (n-1),\
         \bm b^1 + \bm b^2 \gleq \bm 1 \bigg\}. \label{eq:Bset}
    \end{align}
    Then both of the following statements hold:
    \begin{enumerate}
    \item The image of $\Tc$ under $\Sigma$ is equal to $\Bc$, i.e. $\Sigma(\Tc) = \Bc$
    \item The mapping $\Sigma : \Tc \rarr \Bc$ is a bijection.
    \end{enumerate}
\end{lemma}

\vspace{-2em}

\begin{pf}
We prove \emph{1)} by showing first that $\Sigma(\Tc) \subseteq \Bc$, and then $\Bc \subseteq \Sigma(\Tc)$. Any $(S_1,S_2) \in \Tc$ satisfies $|S_1| > 0$, $|S_2| > 0$, $|S_1 \cap S_2| = 0$ as per \eqref{eq:Tdef}. Observe that
\begin{align*}
	|S_1| > 0 &\implies \bm 1^T \bsig(S_1) \geq 1, \\
	|S_2| > 0 &\implies \bm 1^T \bsig(S_2) \geq 1.
\end{align*}
Because $S_1,S_2 \subset \V$ and $|S_1 \cap S_2| = 0$, then $|S_1| < n$. Otherwise if $|S_1| = n$ then either $|S_2| = 0$ or $|S_1 \cap S_2| \neq 0$, which both contradict the definition of $\Tc$. Therefore $|S_1| < n$, and by similar arguments $|S_2| < n$. Observe that
\begin{align*}
	|S_1| < n &\implies \bm 1^T \bsig(S_1) \leq n- 1, \\
	|S_2| < n &\implies \bm 1^T \bsig(S_2) \leq n-1.
\end{align*}
Finally, $|S_1 \cap S_2| = 0$ implies that $j \in S_1 \implies j \notin S_2$ and $j \in S_2 \implies j \notin S_1$ $\forall j \in \{1,\ldots,n\}$. Therefore $\sigma_j(S_1) = 1 \implies \sigma_j(S_2) = 0$ and $\sigma_j(S_2) = 1 \implies \sigma_j(S_1) = 0$.
This implies that
\begin{align*}
	|S_1 \cap S_2| = 0 \implies \bm \sigma(S_1) + \bm \sigma(S_2) \gleq \bm 1.
\end{align*}
Therefore for all $(S_1,S_2) \in \Tc$,
$(\bm \sigma(S_1), \bm \sigma(S_2)) = $ $\Sigma(S_1,S_2)$ satisfies the constraints of the set on the RHS of \eqref{eq:Bset}. This implies that $\Sigma(\Tc) \subseteq \Bc$.

Next, we show $\Bc \subseteq \Sigma(\Tc)$ by showing that for all $(\bm b^1, \bm b^2) \in \Bc$,
there exists an $(S_1,S_2) \in \Tc$ such that $(\bm b^1, \bm b^2) = \Sigma(S_1,S_2)$. Choose any $(\bm b^1, \bm b^2) \in \Bc$ and define sets $(S_1,S_2)$ as follows:
\begin{align}
	b^1_j = 1 &\implies j \in S_1,\ j \in \{1,\ldots,n\}, \nonumber \\
	b^1_j = 0 &\implies j \notin S_1, \nonumber \\
	b^2_j = 1 &\implies j \in S_2, \nonumber\\
	b^2_j = 0 &\implies j \notin S_2. \label{eq:manyimplies}
\end{align}
For the considered sets $(S_1,S_2)$, $1 \leq \bm 1^T \bm b^1$ implies $ |S_1| > 0$ and $1 \leq \bm 1^T \bm b^2$ implies $ |S_2| > 0$. In addition since $\bm b^1 + \bm b^2 \gleq \bm 1$, we have $b^1_j = 1 \implies b^2_j = 0$ and $b^2_j = 1 \implies b^1_j = 0$.
By our choice of $S_1$ and $S_2$, we have $b_j^1 = 1 \implies j \in S_1$, and from previous arguments $b_j^1 = 1 \implies b_j^2 = 0 \implies j \notin S_2$. Similar reasoning can be used to show that $b_j^2 = 1 \implies j \notin S_1$. 
These arguments imply that $|S_1 \cap S_2| = 0$. 
Consequently, $(S_1,S_2)$ satisfies all the constraints of $\Tc$ and is therefore an element of $\Tc$. Clearly, by \eqref{eq:manyimplies} we have $\Sigma(S_1,S_2) = (\bm b^1,\bm b^2)$, which shows that there exists an $(S_1,S_2) \in \Tc$ such that $(\bm b^1, \bm b^2) = \Sigma(S_1,S_2)$. Since this holds for all $(\bm b^1, \bm b^2) \in \Bc$, this implies $\Bc \subseteq \Sigma(\Tc)$. Therefore $\Sigma(\Tc) = \Bc$.

We next prove \emph{2)}. Since $\Sigma(\Tc) = \Bc$, the function $\Sigma : \Tc \rarr \Bc$ is surjective.
To show that it is injective, consider any $\Sigma(S_1,S_2) \in \Bc$ and $\Sigma(\bar{S}_1, \bar{S}_2) \in \Bc$ such that $\Sigma(S_1,S_2) = \Sigma(\bar{S}_1, \bar{S}_2)$. This implies $(\bsig(S_1),\bsig(S_2)) = (\bsig(\bar{S}_1), \bsig(\bar{S}_2))$. Note that $(\bsig(S_1),\bsig(S_2)) = (\bsig(\bar{S}_1), \bsig(\bar{S}_2))$ if and only if $\bsig(S_1) = \bsig(\bar{S}_1)$ and $\bsig(S_2) = \bsig(\bar{S}_2)$. Since the indicator function $\bm \sigma : \Pc(\V) \rarr \{0,1\}^n$ is itself injective, this implies $S_1 = \bar{S}_1$ and $S_2 = \bar{S}_2$, which implies $(S_1,S_2) = (\bar{S}_1,\bar{S}_2)$. Therefore $\Sigma : \Tc \rarr \Bc$ is injective. \hspace*{\fill} $\blacksquare$
\end{pf}

Using Lemma 4 allows us to present the following mixed integer program which solves Problem \ref{prob:FirstProblem}:

\begin{theorem}
\label{thm:rrobust}
    Let $\D$ be an arbitrary nonempty, nontrivial, simple digraph and let $L$ be the Laplacian matrix of $\D$. The maximum $r$-robustness of $\D$, denoted $\rmax(\D)$, is obtained by solving the following minimization problem:
    \begin{align}
            \rmax(\D) =& \hspace{.5em} \underset{\bm b^1,\bm b^2}{\min}
            & & \max \pth{ \max_i \pth{L_i \bm b^1}, \max_j \pth{L_j \bm b^2} } \nonumber \\
            & \hspace{-.5em} \text{\textup{subject to}}
            & & \bm b^1 + \bm b^2 \gleq \bm 1 \nonumber \\
            & & & 1 \leq \bm 1^T \bm b^1 \leq (n-1) \nonumber \\
            & & & 1 \leq \bm 1^T \bm b^2 \leq (n-1) \label{eq:finalprobS0} \nonumber \\
            & & & \bm b^1, \bm b^2 \in \{0,1\}^n.
    \end{align}
    Furthermore, \eqref{eq:finalprobS0} is equivalent to the following mixed integer linear program:
    \begin{align}
            \rmax(\D)=
            \hspace{.5em} \underset{t, \bm b}{\min \hspace{.5em}}
            &\ t \nonumber\\
             \hspace{.5em} \text{\textup{subject to \hspace{.5em}}}
             &0 \leq t,\ t \in \R,\ \bm b \in \Z^{2n} \nonumber\\
             &\bmx{L & \bm 0 \\ \bm 0 & L} \bm b \gleq t \bmx{\bm 1 \\ \bm 1} \nonumber\\
			 & \bm 0 \gleq \bm b \gleq \bm 1 \nonumber \\              
              &\bmx{I_{n\times n} & I_{n \times n}}\bm b \gleq \bm 1 \nonumber\\
              &1 \leq  \bmx{\bm 1^T & \bm 0} \bm b \leq n-1 \nonumber \\
              &1 \leq \bmx{\bm 0 & \bm 1^T} \bm b \leq n-1 \label{eq:probaffine}
\end{align}
\end{theorem}

\vspace{-3em}

\begin{pf}
From Lemmas \ref{lem:rrobalt} and \ref{lem:maxequal}  we have
\begin{align}
            \rmax(\D) = \nonumber \\
             \underset{S_1,S_2 \in \Pc(\V)}{\min}
            & & \max \pth{ \max_i \pth{L_i \bm \sigma(S_1)}, \max_j \pth{L_j \bm \sigma(S_2)} } \nonumber \\
             \text{subject to}
            & & |S_1| > 0,\ |S_2| > 0,\ |S_1 \cap S_2| = 0, \nonumber
    \end{align}
for $i,j \in \{1,\ldots,n\}$. As per Remark \ref{rmk:implicit}, the definition of $\mathcal{T}$ can be used to make the constraints implicit:
\begin{align}
        \label{eq:rrobT}
            &\rmax(\D) = \nonumber \\
            &\underset{(S_1,S_2) \in \mathcal{T}}{\min}
             \max \pth{ \max_i \pth{L_i \bm \sigma(S_1)}, \max_j \pth{L_j \bm \sigma(S_2)} },           
\end{align}
for $i,j \in \{1,\ldots,n\}$. Since $\Sigma: \mathcal{T} \rarr \Bc$ is a bijection by Lemma \ref{lem:bijec}, \eqref{eq:rrobT} is equivalent to
\begin{align}
    \label{eq:implicitB}
            \rmax(\D) =& \underset{(\bm b^1, \bm b^2) \in \Bc}{\min}
            & & \max \pth{ \max_i \pth{L_i \bm b^1}, \max_j \pth{L_j \bm b^2} } \nonumber \\
            & & & i,j \in \{1,\ldots,n\}.
\end{align}
Making the constraints of \eqref{eq:implicitB} explicit yields \eqref{eq:finalprobS0}. 

Next, we prove that \eqref{eq:probaffine} is equivalent to \eqref{eq:finalprobS0}.
The variables $\bm b^1$ and $\bm b^2$ from \eqref{eq:finalprobS0} are combined into the variable $\bm b \in \Z^{2n}$ in \eqref{eq:probaffine}; i.e. $\bm b =\bmxs{(\bm b^1)^T & (\bm b^2)^T}^T $. The first and third constraints of \eqref{eq:probaffine} restrict $\bm b \in \{0,1 \}^{2n}$. Next, it can be demonstrated \cite[Chapter 4]{boyd2004convex} that the formulation ${\min_{\bm x}} \max_i (x_i)$ is equivalent to 
$\min_{t,x} t$ subject to $0 \leq t$, $\bm x \gleq t \bm 1$.    
    Reformulating the objective of the RHS of \eqref{eq:finalprobS0} in this way yields the objective and first two constraints of \eqref{eq:probaffine}:
    \begin{equation}
        \begin{aligned}
            & \underset{t, \bm b}{\min}
            & & t \\
            & \text{subject to}
            &  &0 \leq t,\ \bmx{L & \bm 0 \\ \bm 0 & L} \bm b \gleq t \bmx{\bm 1 \\ \bm 1}. \\
        \end{aligned}
    \end{equation}
    The fourth, fifth, and sixth constraints of \eqref{eq:probaffine} restrict $(\bm b^1, \bm b^2) \in \Bc$ and are simply a reformulation of the first three constraints in \eqref{eq:finalprobS0}. \hspace*{\fill} $\blacksquare$
\end{pf}

\section{Determining $(r,s)$-Robustness using Mixed Integer Linear Programming}
\label{sec:rsrobust}

In this section we address Problems \ref{prob:SecondProblem} and \ref{prob:ThirdProblem}, which involve determining the $(r^*,s^*)$-robustness and $(F_{\max}+1, F_{\max}+1)$-robustness of a given digraph.
To determine these values, we will use the following notation:
\begin{define}
\label{def:smax}
For a digraph $\D$ and a given $r \in \Z_+$, the maximum integer $s$ for which $\D$ is $(r,s)$-robust is denoted as $\smax(r) \in \Z_+$. If $\D$ is not $(r,s)$-robust for any $1 \leq s \leq n$, we will denote $\smax(r) = 0$.
\end{define}
Using this notation, the $(r^*,s^*)$-robustness of a nonempty, nontrivial simple digraph satisfies $r^* = \rmax(\D)$ and $s^* = \smax(\rmax(\D))$. This can be verified by recalling that $r$-robustness is equivalent to $(r,1)$-robustness \cite[Property 5.21]{leblanc2013algorithms}, and that $(r^*,s^*)$ is the maximum element of $\Theta$ according to the lexicographic ordering defined in Section \ref{sec:notation}. Since a method for determining $\rmax(\D)$ has already been presented, this section will introduce a method for determining $\smax(r)$ for any given $r \in \Z_+$. This can then be used to find $\smax(\rmax(\D))$ after $\rmax(\D)$ is determined. 

Recall that $\lor$ indicates logical OR. An equivalent definition of $\smax(r)$ can be given using the following notation:
\begin{define}
\label{def:thetar}
Let $\Theta$ be the set of all $(r,s)$ values for which a given digraph $\D$ is $(r,s)$-robust, as per \eqref{eq:Thetadef}. Let $r \in \Z_+$, and let $\Xr{r}{}$ be defined as in Definition \ref{def:rsrobust}. The set $\Theta_r \subset \Theta$ is defined as follows:
\begin{align}
\label{eq:Thetar}
    \Theta_r = \{ &s \in \Z_+ : \forall (S_1,S_2) \in \Tc,\ \pth{|\Xr{r}{1}| = |S_1|} \lor \nonumber\\
    & \pth{|\Xr{r}{2}| = |S_2|} \lor \pth{|\Xr{r}{1}| + |\Xr{r}{2}| \geq s} \},
\end{align}
\end{define}
In words, $\Theta_r$ is the set of all integers $s$ for which the given digraph $\D$ is $(r,s)$-robust for a given $r \in \Z_+$.
By this definition, $\smax(r) = \max \Thr$, i.e. $\smax(r)$ is simply the maximum element of $\Thr$.

\begin{figure}
\centering
\includegraphics[width=\columnwidth]{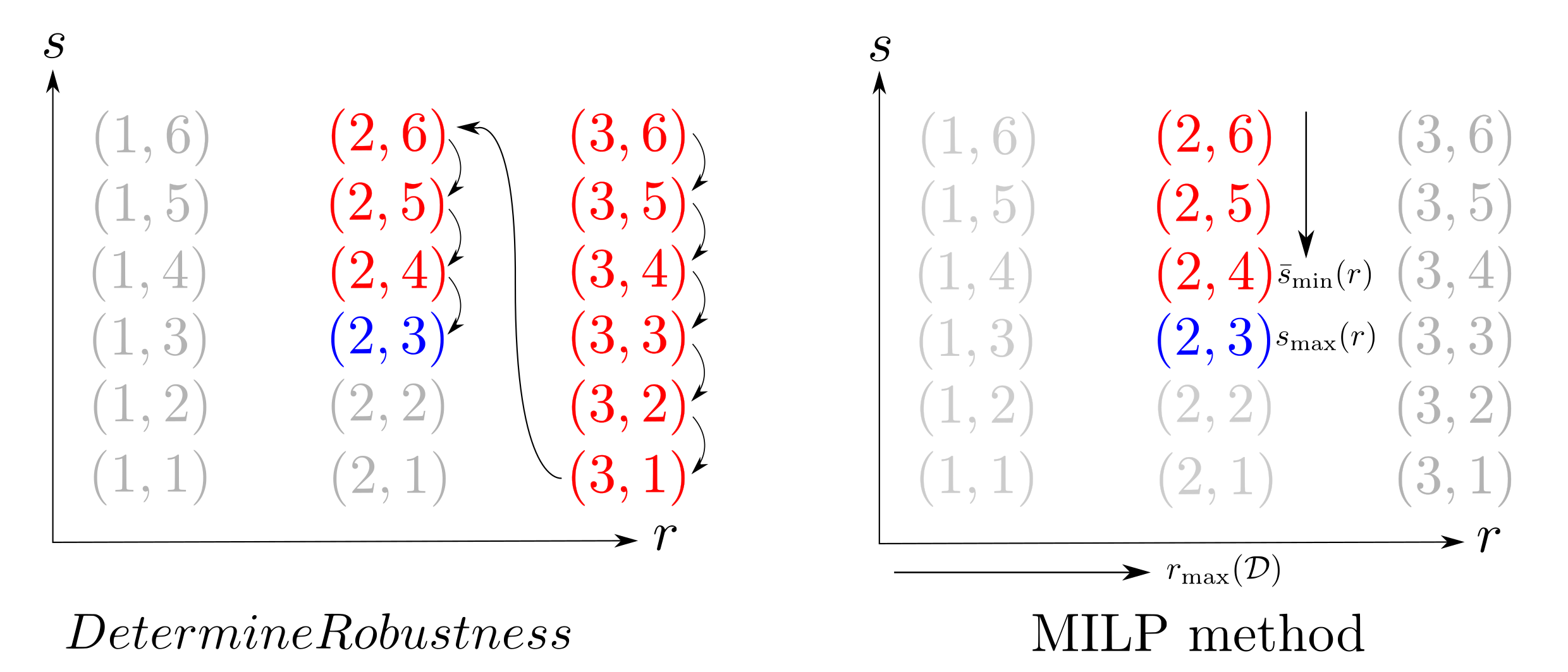}
\caption{
{\scriptsize
Illustration of how the $(r^*,s^*)$-robustness of a graph is found by the $DetermineRobustness$ algorithm and the MILP method. Consider a digraph $\D$ of $n = 6$ nodes which satisfies $(r^*,s^*) = (2,3)$. $DetermineRobustness$ begins with the maximum possible $r$ and $s$ values ($r = \lceil n/2 \rceil$ and $s = n$), then iterates in a lexicographically decreasing manner. The MILP formulation first determines $\rmax(\D)$, then $\bar{s}_{\min}(\rmax(\D))$, then finally infers $\smax(\rmax(D))$ (abbreviated to $\bar{s}_{\min}(r)$ and $\smax(r)$ for clarity).
}
}
\label{fig:DetvsMILP}
\end{figure}

As per \eqref{eq:Thetar}, checking directly if an integer $s \in \Theta_r$ involves testing a logical disjunction for \emph{all} possible $(S_1,S_2)\in \Tc$. This quickly becomes impractical for large $n$ since $|\Tc|$ grows exponentially with $n$. This difficulty can be circumvented, however, by defining the set
\begin{align}
\label{eq:bartheta}
    \bar{\Theta}_r &= \Z_+ \backslash \Theta_r \nonumber \\
    &= \{\bar{s} \in \Z_+ : \bar{s} \notin \Thr \} \nonumber \\
    &= \{ \bar{s} \in \Z_+ : \exists (S_1,S_2) \in \Tc \text{ s.t. } |\pth{\Xr{r}{1}| < |S_1|} \land \nonumber \\
    & \hspace{2em} \pth{|\Xr{r}{2}| < |S_1|} \land \pth{|\Xr{r}{1}| + |\Xr{r}{2}| < \bar{s}} \}, &
\end{align}
where $\land$ denotes logical AND. The set $\bThr$ contains all integers $\bar{s}$ for which the given digraph is \emph{not} $(r,\bar{s})$-robust for the given value of $r$.
\begin{define}
\label{def:smin}
For a digraph $\D$ and a given $r \in \Z_+$, the minimum integer $\bar{s}$ for which $\D$ is \emph{not} $(r,\bar{s})$-robust is denoted as $\smin(r) \in \Z_+$.
\end{define}
As a simple example, consider a digraph $\D$ of 7 nodes where $\smax(3) = 2$. This implies that $\Theta_3 = \{1,2\}$, i.e. the digraph is $(3,1)$- and $(3,2)$-robust. In this case, the set $\bar{\Theta}_3 = \{3,4,5,\ldots\}$ since $\D$ is not $(3,3)$-robust, $(3,4)$-robust, etc. Here we have $\smin(3) = 3$. In general, observe that by definitions \ref{def:smax} and \ref{def:smin} we have
\vspace{-.5em}
\begin{align}
	\label{eq:smaxandsmin}
	\smax(r) = \smin(r) - 1.
\end{align}
It is therefore sufficient to find $\smin(r)$ in order to determine $\smax(r)$. An illustration is given in Figure \ref{fig:DetvsMILP}. The methods in this section will solve for $\smin(r)$ using a mixed integer linear program.
Note that since possible values of $\smax(r)$ are limited to $0 \leq \smax(r) \leq n$ (Definition \ref{def:smax}), possible values of $\smin(r)$ are limited to $1 \leq \smin(r) \leq n+1$.
We point out that it is easier to test if an integer $\bar{s} \in \bThr$ than to test if an integer $s \in \Thr$, in the sense that only one element $(S_1,S_2) \in \Tc$ is required to verify that $\bar{s} \in \bThr$ (as per \eqref{eq:bartheta}) whereas \emph{all} $(S_1,S_2) \in \Tc$ must be checked to verify that $s \in \Thr$ (as per \eqref{eq:Thetar}).

The following Lemma is needed for our main result. It shows
 that given any $r \in \Z_+$ and $S \subset \V$,  the indicator vector of the set $\Xr{r}{}$, denoted $\bm \sigma(\Xr{r}{})$, can be expressed using an MILP. 
Recall that $\Xr{r}{}$ is the set of agents in $S$ which have $r$ in-neighbors outside of $S$, implying
$\sigma_j(\Xr{r}{}) = 1$ if
$L_j \bsig(S) = |\N_j \backslash S| \geq r$.

\begin{lemma}
\label{lem:indicX}
    Let $\D = (\V,\E)$ be an arbitrary nonempty, nontrivial, simple digraph. Let $L$ be the Laplacian matrix of $\D$, and let $r \in \Nat$. Consider any subset $S \subset \V$, $|S| > 0$ and let $\Xr{r}{}$ be defined as in Definition \ref{def:rsrobust}. Then the following holds:
    \vspace{-.5em}
    \begin{align}
    \bm \sigma(\Xr{r}{}) = &\hspace{.5em} \underset{\bm y}{\arg\min}
            & &\bm 1^T \bm y \nonumber\\
            & \hspace{.5em} \text{\textup{subject to}}
            & & L \bm \sigma(S) - (n) \bm y \gleq (r-1) \bm 1 \nonumber \\
            & & & \bm y \in \{0,1\}^n. \label{eq:yind}
\end{align}
\end{lemma}

\vspace{-2em}

\begin{pf}
    Recall that the entries of the indicator vector $\bm \sigma(\Xr{r}{})$ are defined as
\begin{equation}
    \sigma_j(\Xr{r}{}) = \begin{cases}
    1, & \text{if } j \in \Xr{r}{}, \\
    0, & \text{otherwise}.
    \end{cases}
\end{equation} 
Let $\bm y^*$ be an optimal point of the RHS of \eqref{eq:yind}. 
To prove that $\bm y^* = \bm \sigma(\Xr{r}{})$, we demonstrate that $\forall j \in \{1,\ldots,n\}$, $\sigma_j(\Xr{r}{}) = 1 \iff y^*_j = 1$. 
Observe that this is equivalent to demonstrating $\sigma_j(\Xr{r}{}) \neq 1 \iff y^*_j \neq 1\ \forall j$, which is equivalent to demonstrating $\sigma_j(\Xr{r}{}) = 0 \iff y^*_j = 0\ \forall j$. This can be seen by noting $\bm \sigma(\Xr{r}{}) \in \{0,1\}^n$ which implies $\sigma_j(\Xr{r}{}) \neq 1 \iff \sigma_j(\Xr{r}{}) = 0$, and $\bm y^* \in \{0,1\}^n$ which implies $ y^*_j \neq 1 \iff  y^*_j = 0$. Since proving $\sigma_j(\Xr{r}{}) = 1 \iff y^*_j = 1$ for all $j \in \{1,\ldots,n\}$ is equivalent to proving $\sigma_j(\Xr{r}{}) = 0 \iff y^*_j = 0$ for all $j \in \{1,\ldots,n\}$, and both $\bm y^* \in \{0,1\}^n$ and $\bm \sigma(\Xr{r}{}) \in \{0,1\}^n$, we therefore have $(\sigma_j(\Xr{r}{}) = 1 \iff y^*_j = 1\ \forall j)$ if and only if $(\bm y^* = \bm \sigma(\Xr{r}{}))$.



\vspace{-1em}

\emph{Sufficiency:} Consider any $j \in \{1,\ldots,n\}$ such that $\sigma_j(\Xr{r}{}) = 1$. This implies $j \in \Xr{r}{}$ and therefore $|\N_j \backslash S| \geq r$ by Definition \ref{def:rsrobust}. By Lemma \ref{lem:Ljsig}, $|\N_j \backslash S| = L_j \bm \sigma(S)$, and therefore $L_j \bm \sigma(S) > (r-1)$. Since $\bm y^*$ is an optimal point, it is therefore a feasible point. If $y^*_j = 0$, the $j$th row of the first constraint on the RHS of \eqref{eq:yind} is be violated since $L_j \bm \sigma(S) - (n) y^*_j = L_j \bm \sigma(S) \nleq (r-1)$. Therefore we must have $y^*_j = 1$. Note that $|\N_j \backslash S| \leq n\ \forall j \in S$ for any $S \subset \V$.

\vspace{-1em}

\emph{Necessity:} We prove by contradiction. Suppose $y^*_j = 1$ and $\sigma_j(\Xr{r}{}) = 0$. This implies that $ L_j \bm \sigma(S) = |\N_j \backslash S| < r$. Consider the vector $\tilde{\bm y}$ where $\tilde{y}_j = 0$ and $\tilde{y}_i = y^*_i\ \forall i \neq j,\ i \in \{1,\ldots,n\}$. Since $ L_j \bm \sigma(S) = |\N_j \backslash S| < r$, then $\tilde{\bm y}$ is therefore also a feasible point, and $\bm 1^T \tilde{\bm y} < \bm 1^T \bm y^*$. This contradicts $\bm y^*$ being an optimal point to \eqref{eq:yind}; therefore we must have $\sigma_j(\Xr{r}{}) = 1$. \hspace*{\fill} $\blacksquare$
\end{pf}

\vspace{-1em}

The next Theorem presents a mixed integer linear program which determines $\smin(r)$ for any fixed $r \in \Z_+$.

\begin{theorem}
\label{thm:rsgnl}
Let $\D = (\V,\E)$ be an arbitrary nonempty, nontrivial, simple digraph. Let $L$ be the Laplacian matrix of $\D$. Let $r \in \Z_+$, and let $\smin(r)$ be the minimum value of $s$ for which $\D$ is \emph{not} $(r,s)$-robust. Then if $\smin(r) < n+1$, the following holds:
\begin{align}
\label{eq:sbarmax}
            &\smin(r)= & & \nonumber \\
            &\hspace{.5em} \underset{\bar{s}, \bm b^1, \bm b^2, \bm y^1, \bm y^2}{\min}
            &\bar{s} \nonumber\\
            & \hspace{.5em} \text{\textup{subject to}}
            & &1 \leq \bar{s} \leq n + 1,\ \bar{s} \in \Z \nonumber\\
            & & &\bm 1^T \bm y^1 \leq \bm 1^T \bm b^1 - 1 \nonumber \\
            & & &\bm 1^T \bm y^2 \leq \bm 1^T \bm b^2 -1 \nonumber\\
            & & &\bm 1^T \bm y^1 + \bm 1^T \bm y^2 \leq (\bar{s} - 1) \nonumber\\
            & & &\bmx{L & 0 \\ 0 & L} \bmx{\bm b^1 \\ \bm b^2} - (n) \bmx{\bm y^1 \\ \bm y^2} \gleq (r-1) \bmx{\bm 1 \\ \bm 1} \nonumber \\
            & & & \bm b^1 + \bm b^2 \gleq \bm 1 \nonumber \\
            & & & 1 \leq \bm 1^T \bm b^1 \leq (n-1) \nonumber \\
            & & & 1 \leq \bm 1^T \bm b^2 \leq (n-1) \nonumber \\
            & & & \bm b^1, \bm b^2, \bm y^1, \bm y^2 \in \{0,1\}^n.
\end{align}
Furthermore, for any $r > 0$, $\smin(r) = n+1$ if and only if the integer program in \eqref{eq:sbarmax} is infeasible.
\end{theorem}

\vspace{-2em}

\begin{pf}
Note that the theorem statement assumes $r$ is a fixed integer. First, consider the case where $\smin(r) < (n+1)$.
The value of $\smin$ can be found by solving the problem $\smin(r) = \min_{\bar{s} \in \bThr} \bar{s}$.
Making the constraints explicit yields
\begin{align}
    \smin(r,\D)=
            &\hspace{.5em} \underset{\bar{s}, (S_1,S_2) \in \Tc}{\min}
            &\bar{s} \nonumber\\
            & \hspace{.5em} \text{subject to}
            & & |\Xr{r}{1}| \leq |S_1| - 1 \nonumber \\
            & & & |\Xr{r}{2}| \leq |S_2| - 1 \nonumber \\
            & & & |\Xr{r}{1}| + |\Xr{r}{2}| \leq \bar{s} - 1. \label{eq:firststepopt}
\end{align}
To put this problem in an MILP form,
we show that the terms $|\Xr{r}{1}|$, $|\Xr{r}{2}|$, $|S_1|$, and $|S_2|$ can be represented by functions of binary vectors.
This can be done by first observing that for any $S \subseteq \V$, $|S| = \bm 1^T \bm \sigma(S)$. Therefore the following relationships hold:
\begin{align*}
	|S_1| &= \bm 1^T \bm \sigma(S_1), &|\Xr{r}{1}| &= \bm 1^T \bm \sigma(\Xr{r}{1}), \\
	|S_2| &= \bm 1^T \bm \sigma(S_2), &|\Xr{r}{2}| &= \bm 1^T \bm \sigma(\Xr{r}{2}).
\end{align*}
Equation \eqref{eq:firststepopt} can therefore be rewritten as
\begin{align}
    \smin(r,\D)=
            &\hspace{.5em} \underset{\bar{s}, (S_1,S_2) \in \Tc}{\min}
            & \hspace{-1em} \bar{s} \nonumber\\
            & \hspace{-3em} \text{subject to}
            & & \hspace{-4em} \bm 1^T \bm \sigma(\Xr{r}{1}) \leq \bm 1^T \bm \sigma(S_1) - 1 \nonumber \\
            & & & \hspace{-4em} \bm 1^T \bm \sigma(\Xr{r}{2}) \leq \bm 1^T \bm \sigma(S_2) - 1 \nonumber \\
            & & & \hspace{-4em} \bm 1^T \bm \sigma(\Xr{r}{1}) + \bm 1^T \bm \sigma(\Xr{r}{2}) \leq \bar{s} - 1. \label{eq:expandedsbar}
\end{align}
By Lemma \ref{lem:bijec},
the terms $\bm \sigma(S_1),\ \bm \sigma(S_2)$ for $(S_1,S_2) \in \Tc$ can be represented by vectors $(\bm b^1,\bm b^2) \in \Bc$. This yields
\begin{align}
    \smin(r,\D)=
            &\hspace{.5em} \underset{\bar{s}, \bm b^1, \bm b^2}{\min}
            &\hspace{-1.5em} \bar{s} \nonumber\\
            & \hspace{-1em} \text{subject to}
            & &  \bm 1^T \bm \sigma(\Xr{r}{1}) \leq \bm 1^T \bm  b^1 - 1 \nonumber \\
            & & & \bm 1^T \bm \sigma(\Xr{r}{2}) \leq \bm 1^T \bm b^2 - 1 \nonumber \\
            & & &  \bm 1^T \bm \sigma(\Xr{r}{1}) + \bm 1^T \bm \sigma(\Xr{r}{2}) \leq \bar{s} - 1 \nonumber \\
            & & & (\bm b^1 , \bm b^2) \in \Bc. \label{eq:furtherexpandedsbar}
\end{align}
Expanding the last constraint using the definition of $\Bc$ in \eqref{eq:Bset} yields the sixth, seventh, and eighth constraints in \eqref{eq:sbarmax} as well as the constraint that $\bm b^1, \bm b^2 \in \{0,1\}^n$. In addition, the first constraint of the RHS of \eqref{eq:sbarmax} limits the search for feasible value of $\bar{s}$ to the range of possible values for $\smin(r)$.

The vectors $\bm y^1, \bm y^2$ are constrained to satisfy $\bm y^1 = \bm \sigma(\Xr{r}{1})$ and $\bm y^2 = \bm \sigma(\Xr{r}{2})$ as follows: by Lemma \ref{lem:bijec}, $\bm b^1 = \bm \sigma(S_1)$ and $\bm b^2 = \bm \sigma(S_2)$ for $(S_1,S_2) \in \Tc$ as per the sixth through ninth constraints. Therefore by Lemma \ref{lem:indicX},
\begin{align}
    \bm \sigma(\Xr{r}{1}) = &\hspace{.5em} \underset{\bm y^1}{\arg\min}
            & &\bm 1^T \bm y^1 \nonumber\\
            & \hspace{.5em} \text{subject to}
            & & L \bm b^1 - (n) \bm y^1 \gleq (r-1) \bm 1 \nonumber \\
            & & & \bm y^1 \in \{0,1\}^n, \label{eq:S1sigmaOpt} \\
    \bm \sigma(\Xr{r}{2}) = &\hspace{.5em} \underset{\bm y^2}{\arg\min}
            & &\bm 1^T \bm y^2 \nonumber\\
            & \hspace{.5em} \text{subject to}
            & & L \bm b^2 - (n) \bm y^2 \gleq (r-1) \bm 1 \nonumber \\
            & & & \bm y^2 \in \{0,1\}^n. \label{eq:S2sigmaOpt}
\end{align}
The constraints of \eqref{eq:S1sigmaOpt} and \eqref{eq:S2sigmaOpt} are contained in the fifth and last constraints of \eqref{eq:sbarmax}.
Since the fourth constraint of \eqref{eq:sbarmax}, $\bm 1^T \bm y^1 + \bm 1^T \bm y^2 \leq (\bar{s} -1)$, simultaneously minimizes $\bm 1^T \bm y^1$ and $\bm 1^T \bm y^2$, the fourth, fifth, and last constraints of \eqref{eq:sbarmax} ensure that $\bm y^1 = \bm \sigma(\Xr{r}{1})$ and $\bm y^2 = \bm \sigma(\Xr{r}{2})$. Therefore $\bm 1^T \bm y^1 = |\Xr{r}{1}|$ and $\bm 1^T \bm y^2 = |\Xr{r}{2}|$.

Now, by the above arguments,
solving the RHS of \eqref{eq:sbarmax} yields $\smin(r)$ when $\smin(r) < (n+1)$.
We now prove that for $r > 0$, $\smin(r) = n+1$ if and only if the RHS of \eqref{eq:sbarmax} is infeasible. Note that if $r = 0$, then it trivially holds by Definition \ref{def:rsrobust} that $\smax(0) = n$ and therefore $\smin(0) = n+1$.

\emph{Sufficiency:} $\smin(r) = n+1$ implies that $\smax(r) = n$. Recall that $\D$ is $(r, \smax(r))$-robust by Definition \ref{def:smax}, since $\smax(r)$ is the largest integer $s$ for which $\D$ is $(r,s)$-robust. By Definition \ref{def:rsrobust}, this implies that for all $(S_1,S_2) \in \Tc$, at least one of the following three conditions holds: $|\Xr{r}{1}| = |S_1|$, or $|\Xr{r}{2}| = |S_2|$, or $|\Xr{r}{1}| + |\Xr{r}{2}| \geq \smax(r) = n$. Given any $(S_1,S_2) \in \Tc$, we consider each condition separately and show that at least one constraint of \eqref{eq:sbarmax} is violated if the condition holds true:
\begin{itemize}
    \item $\condA$ being true implies that the second constraint of \eqref{eq:sbarmax} is violated. This can be shown using earlier arguments from this proof. Specifically, we have $\bm 1^T \bm y^1 = \condA = \bm 1^T \bm b^1 > \bm 1^T \bm b^1 - 1$. Therefore no feasible point can be constructed from the given set pair $(S_1,S_2)$ if $\condA$.
    \item $\condB$ being true implies that the third constraint of \eqref{eq:sbarmax} is violated. Specifically, we have $\bm 1^T \bm b^2 = |S_2|$ and $\bm 1^T \bm y^2 = |\Xr{r}{2}|$. This implies that $\bm 1^T \bm y^2 = \condB = \bm 1^T \bm b^2 > \bm 1^T \bm b^2 - 1$. Therefore no feasible point can be constructed from the given set pair $(S_1,S_2)$ if $\condB$.
    \item $\condC{n}$ being true implies that 
    both $|\Xr{r}{1}| = |S_1|$ and $\Xr{r}{2} = |S_2|$.
    This follows by observing that $\Xr{r}{1} \subseteq S_1$ and $\Xr{r}{2} \subseteq S_2$, $S_1 \cap S_2 = \{\emptyset\}$ by definition of $\Tc$ in \eqref{eq:Tdef}, and therefore $\Xr{r}{1} \cap \Xr{r}{2} = \{\emptyset \}$. 
    Since $S_1,S_2 \subset \V$ and $|\V| = n$,
    we have $n \leq |\Xr{r}{1}| + |\Xr{r}{2}| \leq |S_1| + |S_2| \leq |\V| = n$. We must therefore have $|\Xr{r}{1}| = |S_1|$ and $|\Xr{r}{2}| = |S_2|$,
    which from prior arguments both imply that a constraint of \eqref{eq:sbarmax} is violated. Therefore no feasible point can be constructed from the given set pair $(S_1,S_2)$ if $\condC{n}$
\end{itemize}
Since for all $(S_1,S_2) \in \Tc$ at least one of these three conditions holds, for all $(S_1,S_2) \in \Tc$ at least one constraint of \eqref{eq:sbarmax} is violated when $\smax(r) = n$, which is equivalent to $\smin(r) = n+1$.
Therefore $\smin(r) = n + 1$ implies that \eqref{eq:sbarmax} is infeasible.

\emph{Necessity:} We prove the contrapositive, i.e. we prove that $\smin \neq n+1$ implies that there exists a feasible point to the RHS of \eqref{eq:sbarmax}. First, no digraph on $n$ nodes is $(r,n+1)$-robust \cite[Definition 13]{leblanc2013resilient}, and the contrapositive of Property \ref{prop:rsimplies} implies that if a graph is \emph{not} $(\bar{r},\bar{s})$-robust, then it is also not $(\bar{r}',\bar{s}')$-robust for all $\bar{r}' \geq \bar{r}$, and for all $\bar{s}' \geq \bar{s}$. Therefore $\smin \neq n+1$ implies $\smin \leq n$.
Next, $\smin \leq n$ implies $n \in \bThr$, which implies that there exists $(S_1, S_2) \in \Tc$ such that $|\Xr{r}{1}| \leq |S_1| - 1$ and $|\Xr{r}{2}| \leq |S_2| - 1$ and $|\Xr{r}{1}| + |\Xr{r}{2}| \leq n-1$, as per \eqref{eq:bartheta}. Letting $\bar{s} = n$, $\bm b^1 = \bm \sigma(S_1)$, $\bm b^2 = \bm \sigma(S_2)$, $\bm y^1 = \bm \sigma(\Xr{r}{1})$, and $\bm y^2 = \bm \sigma(\Xr{r}{2})$ yields a feasible point to \eqref{eq:sbarmax}. \hspace*{\fill} $\blacksquare$
\end{pf}
The MILPs in Theorems \ref{thm:rrobust} and \ref{thm:rsgnl} can be used to determine the $(r^*,s^*)$-robustness of any digraph satisfying Assumption \ref{assume:simple}, thereby solving Problem \ref{prob:SecondProblem}. Recall from the beginning of Section \ref{sec:rsrobust} that $r^* = \rmax(\D)$ and $s^* = \smax(\rmax(\D))$. Theorem \ref{thm:rrobust} can first be used to determine the value of $\rmax(\D) = r^*$. Using $\rmax(\D)$, Theorem \ref{thm:rsgnl} can then be used to find the value of $\smax(\rmax(\D)) = s^*$.

More generally however, the MILP formulation in Theorem \ref{thm:rsgnl} allows for $\smax(r)$ to be determined for \emph{any} $r \in \Z_+$. Since $(r,1)$-robustness is equivalent to $r$-robustness, the MILP in Theorem \ref{thm:rsgnl} can also be used to determine whether a digraph $\D$ is $r$ robust for a given $r \in \Z_+$. If $\smin(r) \geq 2$, then $\smax(r) \geq 1$ which implies that $\D$ is $(r,1)$-robust. On the other hand, $\smin(r) = 1$ implies that $1 \in \bThr$ and therefore $\D$ is \emph{not} $(r,1)$-robust (and \emph{not} $r$-robust).

Finally, to solve Problem \ref{prob:ThirdProblem} Theorem \ref{thm:rsgnl} can be used to determine the $(F_{\max}+1, F_{\max}+1)$-robustness of a nonempty, nontrivial, simple digraph. Recall that $F_{\max} = \max (\{F \in \Z_+ : (F+1,F+1) \in \Theta\})$. The value of $F_{\max}$ is determined by Algorithm \ref{alg:Fmax}, presented below. 
\begin{algorithm}
\caption{\small{\textsc{DetermineFmax}}}
\label{alg:Fmax}
\begin{algorithmic}[1]
\State $r' \larr \rmax(\D)$ from MILP in Theorem \ref{thm:rrobust}
	\While {$r' > 0$}
	\State $s' \larr \smax(r')$ from MILP in Theorem \ref{thm:rsgnl}
	\If {$s' \geq r'$} \Comment{\parbox[t]{.5\linewidth}{\emph{Prop. \ref{prop:rsimplies} implies graph is $(r',s)$-rob. $\forall s \leq s'$, therefore $\D$ is $(r',r')$-robust}}}
		\State $F_{\max} \larr (r' - 1)$
		\State \textbf{return} $F_{\max}$
	\Else
		\State $r' \larr (r' - 1)$
	\EndIf
	\EndWhile
	\State $F_{\max} \larr 0$
	\State \textbf{return} $F_{\max}$
\end{algorithmic}
\end{algorithm}
In essence, Algorithm \ref{alg:Fmax} finds the largest values of $r'$ and $s'$ such that $r' = s'$ and $(r',s') \in \Theta$. It begins by setting $r' \larr \rmax(\D)$, and finding $\smax(r')$ using Theorem \ref{thm:rsgnl}. If $\smax(r') \geq r'$, then by Proposition \ref{prop:rsimplies} the digraph $\D$ is $(r',s)$-robust for $s = r'$ and therefore $(r',r')$-robust. This implies $r' = F_{\max} + 1$. However, if $\smax(r') < r'$ then $r'$ is decremented, $\smax(r')$ recalculated, and the process is repeated until the algorithm terminates with the highest integer $r'$ such that $\D$ is $(r',r')$-robust, yielding $F_{\max} = (r'-1)$.

\vspace{-1em}
\section{Reducing Worst-Case Performance of Determining $\bf \rmax(\D)$}
\label{sec:reducecomplex}

When solving a MILP with zero-one integer variables using a branch-and-bound technique, the worst-case number of subproblems to be solved is equal to $2^q$, where $q$ is the dimension of the zero-one integer vector variable. 
In Section \ref{sec:rrobust}, the MILP in Theorem 1 which solves for $\rmax(\D)$ has a binary vector variable with dimension $2n$. In this section, we present two MILPs whose optimal values provide upper and lower bounds on the value of $\rmax(\D)$. Each MILP has a binary vector variable with dimension of only $n$, which implies a reduced worst-case performance as compared to the MILP in Theorem \ref{thm:rrobust}.


\vspace{-.5em}

\subsection{A Lower Bound on Maximum $r$-Robustness}

\vspace{-.5em}

In \cite{Guerrero2016formations}, a technique is presented for lower bounding $\rmax(\D)$ of undirected graphs by searching for the minimum reachability of subsets $S \subset \V$ such that $|S| \leq \floor{n/2}$. We extend this result to digraphs in the next Lemma.

\begin{lemma}
    \label{lem:rholb}
    Let $\D = (\V,\E)$ be an arbitrary nonempty, nontrivial, simple digraph. Let $\Psi = \{S \subset \V : 1 \leq |S| \leq \floor{n/2} \}$. Then the following holds:
    \begin{align}
    \label{eq:lowerbound}
        \rmax(\D) \geq \min_{S \in \Psi} \Rc(S).
    \end{align}
\end{lemma}
\vspace{-2em}
\begin{pf}
    By Lemma \ref{lem:rrobalt} and Remark \ref{rmk:implicit}, proving \eqref{eq:lowerbound} is equivalent to proving
    \begin{align}
        \min_{S \in \Psi} \Rc(S) \leq \min_{(S_1,S_2) \in \Tc} \max\pth{\Rc(S_1),\Rc(S_2)}.
    \end{align}
    Denote $S^* = \arg \min_{S \in \Psi} \Rc(S)$ and $(S_1^*, S_2^*) = \arg \min_{(S_1,S_2) \in \Tc} \max(\Rc(S_1),\Rc(S_2))$. We prove by contradiction. Suppose $\Rc(S^*) > \max(\Rc(S_1^*),\Rc(S_2^*))$. Since $S_1^*$ and $S_2^*$ are nonempty, $|S_1^*| \geq 1$ and $|S_2^*| \geq 1$. Since they are disjoint, we must have either $|S_1^*| \leq \floor{n/2}$, or $|S_2^*| \leq \floor{n/2}$, or both $|S_1^*|$ and $|S_2^*|$ less than or equal to $\floor{n/2}$. Therefore either $S_1^* \in \Psi$ or $S_2^* \in \Psi$. This implies that either $\Rc(S_1^*) \geq \Rc(S^*)$ (if $S_1^* \in \Psi$) or $\Rc(S_2^*) \geq \Rc(S^*)$ (if $S_2^* \in \Psi$), since $S^*$ is an optimal point. But this contradicts the assumption that $\Rc(S^*) > \max(\Rc(S_1^*),\Rc(S_2^*))$.
Therefore we must have
    \begin{align}
        \min_{S \in \Psi} \Rc(S) \leq \min_{(S_1,S_2) \in \Tc} \max\pth{\Rc(S_1),\Rc(S_2)} = \rmax(\D),
    \end{align}
which concludes the proof. \hspace*{\fill} $\blacksquare$
\end{pf}

Using this result, a lower bound on $\rmax(\D)$ can be obtained by the following optimization problem:

\begin{theorem}
\label{thm:lowerbound}
    Let $\D$ be an arbitrary nonempty, nontrivial, simple digraph and let $L$ be the Laplacian matrix of $\D$. A lower bound on the maximum integer for which $\D$ is $r$-robust, denoted $\rmax(\D)$, is found as follows:
    \begin{equation}
    \label{eq:rholbthm}
        \begin{aligned}
            \rmax(\D) \geq& \hspace{.5em} \underset{\bm b}{\min}
            & & \max_i \pth{L_i \bm b}  \\
            & \hspace{.5em} \text{\textup{subject to}}
            & & 1 \leq \bm 1^T \bm b \leq \floor{n/2} \\
            & & & \bm b \in \{0,1\}^n.
        \end{aligned}
    \end{equation}
    Furthermore, \eqref{eq:rholbthm} is equivalent to the following mixed integer linear program:
    
    \begin{align}
            \rmax(\D) \geq& \hspace{.5em} \underset{t, \bm b}{\min}
            & & t \nonumber \\
            & \hspace{.5em} \text{\textup{subject to}}
            & & t \geq 0,\ \bm b \in \Z^n \nonumber \\
            & & & L \bm b \gleq t \bm 1 \nonumber \\
            & & & \bm 0 \gleq \bm b \gleq \bm 1 \nonumber \\
            & & & 1 \leq \bm 1^T \bm b \leq \floor{n/2}
             \label{cor:rholb}
    \end{align}
\end{theorem}

\vspace{-2em}

\begin{pf}
	To prove the result we show that the RHS of \eqref{eq:rholbthm} is equivalent to the RHS of \eqref{eq:lowerbound}. By Lemma \ref{lem:Ljsig}, $\Rc(S) = \max_i L_i \bm \sigma(S)$. Therefore \eqref{eq:lowerbound} is equivalent to
\begin{align}
    \label{eq:rmaxDgreater}
        \rmax(\D) \geq \min_{S \in \Psi} \max_i L_i \bm \sigma(S).
\end{align}
Next, we demonstrate that the set
\begin{align}
\label{eq:bpsi}
        \Bc_\Psi = \{\bm b \in \{0,1\}^n : 1 \leq \bm 1^T \bm b \leq \floor{n/2} \}
\end{align}
satisfies $\Bc_\Psi = \bsig(\Psi)$, where $\bsig(\Psi)$ is the image of $\Psi$ under $\bm \sigma : \Pc(\V) \rarr \{0,1\}^n$. Since $1 \leq |S| \leq \floor{n/2}$ for all $S \in \Psi$, then by \eqref{eq:sigmadef} we have $1 \leq \bm 1^T \bsig(S) \leq \floor{n/2}$ for all $S \in \Psi$. Also, $\bsig(S) \in \{0,1\}^n$, and therefore $\bsig(S) \in \Bc_\Psi\ \forall S \in \Psi$, implying that $\bsig(\Psi) \subseteq \Bc$. Next, for any $\bm b \in \Bc_\Psi$, choose the set $S = \bsig^{-1}(\bm b)$ (recall from \ref{subsec:reformulate} that $\bsig^{-1} : \{0,1\}^n \rarr \Pc(\V)$). Then clearly $\bsig(S) = \bsig(\bsig^{-1}(\bm b)) = \bm b$, and therefore $\Bc_\Psi \subseteq \bsig(\Psi)$. Therefore $\Bc_\Psi = \bsig(\Psi)$.

The function $\bm \sigma : \Psi \rarr \Bc_\Psi$ is therefore surjective. Since $\bsig : \Pc(\V) \rarr \{0,1\}^n$ is injective, $\Psi \subset \Pc(\V)$, and $\Bc_\Psi \subset \{0,1\}^n$, then $\bm \sigma : \Psi \rarr \Bc_\Psi$ is also injective and therefore a bijection. This implies that \eqref{eq:rmaxDgreater} is equivalent to
\begin{align}
    \label{eq:rmaxDfinalstep}
        \rmax(\D) \geq \min_{\bm b \in \Bc_\Psi} \max_i L_i \bm b.
\end{align}
Making the constraints of \eqref{eq:rmaxDfinalstep} explicit yields \eqref{eq:rholbthm}. More specifically, since $\Bc_\Psi = \{\bm b \in \{0,1\}^n : 1 \leq \bm 1^T \bm b \leq \floor{n/2} \}$ by \eqref{eq:bpsi}, equation \eqref{eq:rmaxDfinalstep} can be rewritten with explicit constraints on $\bm b$ as follows:
 \begin{align}
            \rmax(\D) \geq& \hspace{.5em} \underset{\bm b}{\min}
            & & \max_i \pth{L_i \bm b} \nonumber  \\
            & \hspace{.5em} \text{\textup{subject to}}
            & & 1 \leq \bm 1^T \bm b \leq \floor{n/2} \nonumber \\
            & & & \bm b \in \{0,1\}^n. \label{eq:runningoutofideas}
    \end{align}
Equation \eqref{eq:runningoutofideas} is the same as \eqref{eq:rholbthm}. 

We next prove that \eqref{cor:rholb} is equivalent to \eqref{eq:rholbthm}.
As per the proof of Theorem \ref{thm:rrobust}, the objective and first two constraints of \eqref{cor:rholb} are a reformulation of the objective of the RHS of \eqref{eq:rholbthm}. The first and third constraint ensure $b$ to be in $\{0,1\}^n$, and the fourth constraint ensures $\bm b \in \Bc_\Psi$. \hspace*{\fill} $\blacksquare$
\end{pf}

\subsection{An Upper Bound on Maximum r-Robustness}

\label{sec:twosetprob}
\vspace{-.5em}

This section will present an MILP whose solution provides an \emph{upper} bound on the value of $\rmax(\D)$, and whose binary vector variable has a dimension of $n$. This will be accomplished by searching a subset $\Tc' \subset \Tc$ which is defined as
\begin{equation}
\label{eq:Tcprime}
\Tc' = \{(S_1,S_2) \in \Tc : S_1 \cup S_2 = \V \}.
\end{equation}
In other words, $\Tc'$ is the set of all possible partitionings of $\V$ into $S_1$ and $S_2$. Considering only elements of $\Tc'$ yields certain properties that allow us to calculate an upper bound on $\rmax(\D)$ using an MILP with only an $n$-dimensional binary vector.

Observe that $|\Tc'| = 2^n - 2$, since neither $\Tc$ nor $\Tc'$ include the cases where $S_1 = \{ \emptyset \}$ or $S_2 = \{ \emptyset \}$.
Similar to the methods discussed earlier, the partitioning of $\V$ into $S_1$ and $S_2$ can be represented by the indicator vectors $\bm \sigma(S_1)$ and $\bm \sigma(S_2)$, respectively.
Note that since $S_1 \cup S_2 = \V$ for all $(S_1,S_2) \in \Tc'$, it can be shown that ${\bm \sigma(S_1) + \bm \sigma(S_2) = \bm 1}\ \forall (S_1,S_2) \in \Tc'$.
These properties allow the following Lemma to be proven:

\begin{lemma}
\label{lem:nodereach}
Let $\D= (\V,\E)$ be an arbitrary nonempty, nontrivial, simple digraph. Let $L$ be the Laplacian matrix of $\D$ and let $L_j$ be the $j$th row of $L$. Let $\Tc'$ be defined as in \eqref{eq:Tcprime}. Then for all $(S_1,S_2) \in \Tc'$, the following holds:
\begin{align}
    L_j \bm \sigma(S_1) &= \begin{cases}
        |\N_j \backslash S_1|, & \text{if } j \in S_1, \\
        -|\N_j \backslash S_2|, & \text{if } j \in S_2.
    \end{cases} \nonumber \\
    L_j \bm \sigma(S_2) &= \begin{cases}
        |\N_j \backslash S_2|, & \text{if } j \in S_2, \\
        -|\N_j \backslash S_1|, & \text{if } j \in S_1.
    \end{cases}
\end{align}
\end{lemma}

\vspace{-2.5em}

\begin{pf}
Lemma \ref{lem:Ljsig} implies that $L_j \bm \sigma(S_1) = |\N_j \backslash S_1|$ if $j \in S_1$, and $L_j \bm \sigma(S_2) = |\N_j \backslash S_2|$ if $j \in S_2$. Since $(S_1,S_2) \in \Tc' \implies \bm \sigma(S_1) + \bm \sigma(S_2) = \bm 1$, we have
\begin{align}
    L_j \bm \sigma(S_1) &= L_j (\bm 1 - \sigma(S_2)) = -L_j \bm \sigma(S_2).
\end{align}
This relation holds because, by the definition of $L$, $\bm 1$ is always in the null space of $L$. 
Therefore for $j \in S_2$ we have $L_j \bm \sigma(S_1) = -L_j \bm \sigma(S_2) = -|\N_j \backslash S_2|$, and for $j \in S_1$ we have $L_j \bm \sigma(S_2) = -L_j \bm \sigma(S_1) = -|\N_j \backslash S_1|$. \hspace*{\fill} $\blacksquare$
\end{pf}

An interesting result of Lemma \ref{lem:nodereach} is that for any subsets $(S_1,S_2) \in \Tc'$, the maximum reachability of the two subsets can be recovered using the infinity norm.

\begin{lemma}
\label{lem:2setinf}
    Let $\D= (\V,\E)$ be an arbitrary nonempty, nontrivial, simple digraph and let $L$ be the Laplacian matrix of $\D$. For all $(S_1,S_2) \in \Tc'$, the following holds:
    \begin{equation}
    \label{eq:2setinf}
      \hspace{-.1em} \nrm{L \bm \sigma(S_1)}_\infty = \nrm{L \bm \sigma(S_2)}_\infty = \max(\Rc(S_1),\Rc(S_2)).
    \end{equation}
\end{lemma}
\vspace{-2.5em}
\begin{pf}
    Denote the nodes in $S_1$ as $\{i_1, \ldots,i_p \}$ and the nodes in $S_2$ as $\{j_1,\ldots,j_{(n-p)} \}$ with $p \in \Z,\ 1 \leq p \leq (n-1)$. Note that since $S_1 \cup S_2 = \V$, we have $\{i_1, \ldots, i_p\} \cup \{j_1,\ldots, j_{n-p} \} = \{1,\ldots,n\}$.

    The right hand side of equation \eqref{eq:2setinf} can be expressed as
    \begin{align}
        \max(\Rc(S_1),\Rc(S_2)) =& \hspace{.5em} \max (|N_{i_1} \backslash S_1|,\ldots, |N_{i_p} \backslash S_1|, \nonumber \\
    & \hspace{-.5em} |N_{j_1} \backslash S_2|, \ldots, |N_{j_{(n-p)}} \backslash S_2| ).
    \end{align}
    Similarly, using Lemma \ref{lem:nodereach} yields
    \begin{align}
        \nrm{L \bm \sigma(S_1)}_\infty = \max \big(& |L_1 \bm \sigma(S_1)|,\ldots,|L_n \bm \sigma(S_1)| \big)\nonumber \\
        = \max \big(& |N_{i_1} \backslash S_1|,\ldots, |N_{i_p} \backslash S_1|, \nonumber \\
        &|N_{j_1} \backslash S_2|, \ldots, |N_{j_{(n-p)}} \backslash S_2| \big) \nonumber \\
        &= \max(\Rc(S_1),\Rc(S_2)).
    \end{align}
    Finally, observe that
    \begin{align}
        \nrm{L \bm \sigma(S_2)}_\infty &= \nrm{L \bm (\bm 1 - \bm \sigma(S_1))}_\infty = \nrm{L \bm \sigma(S_1)}_\infty,
    \end{align}
    which completes the proof. \hspace*{\fill} $\blacksquare$
\end{pf}

A mixed integer linear program yielding an upper bound on the value of $\rmax(\D)$ is therefore given by the following Theorem:

\begin{theorem}
\label{thm:twocase}
    Let $\D = (\V,\E)$ be an arbitrary nonempty, nontrivial, simple digraph. Let $L$ be the Laplacian matrix of $\D$. The maximum integer for which $\D$ is $r$-robust, denoted $\rmax(\D)$, is upper bounded as follows:
    \begin{align}
            \rmax(\D) \leq \hspace{.5em}&\underset{\bm b}{\min}
            & & \nrm{L \bm b}_\infty \nonumber\\
            & \text{\textup{subject to}}
            & &1 \leq \bm 1^T \bm b \leq (n-1)  \nonumber \\
            & & &\bm b \in \{0,1\}^n.  \label{eq:thmtwocase}
    \end{align}
    Furthermore, \eqref{eq:thmtwocase} is equivalent to the following mixed integer linear program:
    \begin{align}
\label{eq:infnorm}
            \rmax(\D) \leq& \hspace{.5em} \underset{t, \bm b}{\min}
            & & t \\
            & \hspace{.5em} \text{\textup{subject to}}
            & & 0 \leq t,\ \bm b \in \Z^n. \nonumber \\
            & & & -t \bm 1 \gleq L \bm b \gleq \bm 1 t \nonumber  \\
            & & & \bm 0 \gleq \bm b \gleq \bm 1 \nonumber  \\
            & & & 1 \leq \bm 1^T \bm b \leq (n-1) \nonumber 
\end{align}
\end{theorem}
\vspace{-3em}
\begin{pf}
Consider the optimization problem
\begin{equation}
    \label{eq:rsmalldomain}
        \begin{aligned}
            & \underset{(S_1,S_2) \in \Tc^{'}}{\min}
            & & \max \pth{\Rc(S_1),\Rc(S_2)}.
        \end{aligned}
    \end{equation}
Since $\Tc' \subset \Tc$, the optimal value of \eqref{eq:rsmalldomain} is a valid upper bound on the value of $\rmax(\D)$ as per Remark \ref{rmk:implicit}.
From \eqref{eq:rsmalldomain} and Lemma \ref{lem:2setinf} we obtain
\begin{align}
    \rmax(\D) &\leq \underset{(S_1,S_2) \in \Tc^{'}}{\min} \nrm{L \bm \sigma(S_1)}_\infty \nonumber \\
    &= \underset{(S_1,S_2) \in \Tc^{'}}{\min} \nrm{L \bm \sigma(S_2)}_\infty. \label{eq:rmaxUBinfnorm}
\end{align}
Since $S_1,S_2$ are nonempty and $S_1 \cup S_2 = \V$ for all $ (S_1,S_2) \in \Tc'$, the set of all possible $S_1$ subsets within elements of $\Tc'$ is $(\Pc(\V) \backslash \{\emptyset,\V \})$. Similarly, the set of all possible $S_2$ subsets within elements of $\Tc'$ is also $(\Pc(\V) \backslash \{\emptyset,\V \})$. For brevity, denote $\Pc_{\emptyset,\V} = \Pc(\V) \backslash \{\emptyset, \V\}$.

Next, we demonstrate that the set $\Bc' = \{\bm b \in \{0,1\}^n : 1 \leq \bm 1^T \bm b \leq (n-1)\}$ satisfies $\bsig(\Pce) = \Bc'$. Since $1 \leq |S| \leq (n-1)$ for all $S \in \Pce$, then by \eqref{eq:sigmadef} we have $1 \leq \bm 1^T \bsig(S) \leq (n-1)$ for all $S \in \Pce$. Also, $\bsig(S) \in \{0,1\}^n$, and therefore $\bsig(S) \in \Bc'\ \forall S \in \Pce$, implying that $\bsig(\Pce) \subseteq \Bc'$. Next, for any $\bm b \in \Bc'$, choose the set $S = \bm \sigma^{-1}(\bm b)$. Then clearly $\bsig(S) = \bsig(\bsig^{-1}(\bm b)) = \bm b$, and therefore $\Bc' \subseteq \bsig(\Pce)$. Therefore $\Bc' = \bsig(\Pce)$.

The function $\bsig : \Pce \rarr \Bc'$ is therefore surjective. Since $\bsig : \Pc(\V) \rarr \{0,1\}^n$ is injective, $\Pce \subset \Pc(\V)$, and $\Bc' \subset \{0,1\}^n$, then $\bsig : \Pce \rarr \Bc'$ is also injective. Therefore $\bsig : \Pce \rarr \Bc'$ is a bijection, implying that \eqref{eq:rmaxUBinfnorm} is equivalent to
\begin{align}
	\rmax(\D) &\leq \underset{\bm b \in \Bc'}{\min} \nrm{L \bm b}_\infty. \label{eq:rmaxLBfinal}
\end{align}
Making the constraints of \eqref{eq:rmaxLBfinal} explicit yields \eqref{eq:thmtwocase}.
We next prove that \eqref{eq:infnorm} is equivalent to \eqref{eq:thmtwocase}. 
It can be shown \cite[Chapter 4]{boyd2004convex} that $\min_{\bm x} \nrm{\bm x}_\infty$ is equivalent to 
\begin{equation*}
        \begin{aligned}
            & \underset{t, \bm x}{\min}
            & & t \\
            & \text{subject to}
            & & -t \bm 1 \gleq \bm x \gleq t \bm 1. \\
        \end{aligned}
\end{equation*}
Likewise, the objective and first two constraints of the RHS of \eqref{eq:infnorm} are a reformulation of the objective of the RHS of \eqref{eq:thmtwocase}. The first and third constraint restrict $\bm b \in \{0,1\}^n$, and the fourth constraint restricts $\bm b$ to be an element of $\Bc'$. These arguments imply that the RHS of \eqref{eq:infnorm} is equivalent to the RHS of \eqref{eq:thmtwocase}. \hspace*{\fill} $\blacksquare$
\end{pf}

\section{Discussion}
\label{sec:discussion}

MILP problems are $NP$-hard problems to solve in general. As such, the formulations presented in this paper do not reduce the theoretical complexity of the robustness determination problem. However, it has been pointed out that algorithmic advances and improvement in computer hardware have led to a speedup factor of 800 billion for mixed integer optimization problems during the last 25 years \cite{bertsimas2017optimal}. The results of this paper allow for the robustness determination problem to benefit from ongoing and future improvements in the active areas of optimization and integer programming.

In addition, one crucial advantage of the MILP formulations is the ability to iteratively tighten a global lower bound on the optimal value over time by using a branch-and-bound algorithm. The reader is referred to \cite{wolsey2007mixed} for a concise overview of how such a lower bound can be calculated. In context of robustness determination, lower bounds on $\rmax(\D)$ and $\smax(r)$ are generally more useful than upper bounds since they can be used to calculate lower bounds on the maximum adversary model that the network can tolerate. The ability to use branch-and-bound algorithms for solving the robustness determination problem offers the flexibility of terminating the search for $\rmax(\D)$ and/or $\smax(r)$ when sufficiently high lower bounds have been determined. In this manner, approximations of these values can be found when it is too computationally expensive to solve for them exactly. The investigation of additional methods to find approximate solutions to the MILP formulations in this paper is left for future work.

\vspace{-1em}

\section{Simulations}
\label{sec:sim}

This section presents simulations which demonstrate the performance of the MILP formulations as compared to a robustness determination algorithm from prior literature called $DetermineRobustness$ \cite{leblanc2013algorithms}. Computations for these simulations are performed in MATLAB 2018b on a desktop computer with 8 Intel core i7-7820X CPUs (3.60 GHz) capable of handling 16 total threads. All MILP problems are solved using MATLAB's \emph{intlinprog} function.

Four types of random graphs are considered in the simulations: Erd\H{o}s-R\'enyi random graphs, random digraphs, $k$-out random graphs \cite{bollobas2001random}, and $k$-in random graphs.
The Erd\H{o}s-R\'enyi random graphs in these simulations consist of $n$ agents, with each possible undirected edge present independently with probability $p \in [0,1]$ and absent with probability $1-p$. Three values of $p$ are considered: 0.3, 0.5, and 0.8.
The random digraphs considered consist of $n$ nodes with each possible \emph{directed} edge present independently with probability $p$ and absent with probability $1-p$. The values of $p$ considered are again 0.3, 0.5, and 0.8.
The $k$-out random graphs consist of $n$ nodes. For each vertex $i \in \V$, $k \in \Nat$ other nodes are chosen with all  having equal probability and all choices being independent. For each node $j$ of the $k$ chosen nodes, a directed edge $(i,j)$ is formed.
$k$-in random graphs are formed in the same manner as $k$-out random graphs with the exception that the direction of the directed edges are reversed; i.e. edges $(j,i)$ are formed. The values of $k$ considered are $\{3,4,5\}$.

Two sets of simulations are considered. The first set compares two algorithms which determine the pair $(r^*,s^*)$ for a digraph: the $DetermineRobustness$ algorithm from \cite{leblanc2013algorithms} and Algorithm \ref{alg:MILP}, \emph{$(r,s)$-Rob. MILP}, which is an MILP formulation using results from Theorems \ref{thm:rrobust} and \ref{thm:rsgnl}. Details about the implementation of these algorithms can be found in the Appendix, section \ref{subsec:AppendixAlg}. The algorithms are tested on the four types of graphs described above with values of $n$ ranging from $7$ to $15$. In addition, the MILP formulation is tested on digraphs with values of $n$ ranging from $17$ to $25$. The $DetermineRobustness$ algorithm is not tested on values of $n$ above 15 since the convergence rate trend is clear from the existing data, and the projected convergence times are prohibitive for large $n$. 100 graphs per graph type and combination of $n$ and $p$ (or $n$ and $k$ depending on the respective graph type), are randomly generated, and the algorithms are run on each graph. Overall, 10,800 total graphs are analyzed with $DetermineRobustness$ and 16,800 total graphs are analyzed with Algorithm \ref{alg:MILP}. The time for each algorithm to determine the pair $(r^*,s^*)$ is averaged for each combination of $n$ and $p$ (for Erd\H{o}s-R\'enyi random graphs and random digraphs), and for each combination of $n$ and $k$ (for $k$-out and $k$-in random graphs). 
The interpolated circles represent the average convergence time in seconds over 100 trials for each value of $n$, while the vertical lines represent the spread between maximum convergence time and minimum convergence time over trials for the respective value of $n$. Note the logarithmic scale of the y-axis.


To facilitate the large number of graphs being tested, a time limit of $10^3$ seconds (roughly 17 minutes) is imposed on Algorithm \ref{alg:MILP} (the MILP formulation). However, out of the 16,800 graphs tested by Algorithm \ref{alg:MILP}, there are only 62 instances where the algorithm did not converge to optimality before this time limit. Instances where the time limit was violated are given the maximum time of $10^3$ seconds and included in the data. The graphs where optimality was not reached by the time limit all have between 21 and 25 nodes, are either Erd\H{o}s-R\'enyi random graphs or random digraphs, and have edge formation probabilities of $p = 0.8$.


Several patterns in the data warrant discussion. It is clear that in some cases, the minimum time of $DetermineRobustness$ is less than that of \emph{$(r,s)$-Rob. MILP}. $DetermineRobustness$ terminates if two subsets $S_1$ and $S_2$ which are both 0-reachable are encountered, since this implies that the graph is at most $(0,n)$-robust. This can result in fast termination if such subsets are encountered early in the search. Second, there are instances where the maximum time for the \emph{$(r,s)$-Rob. MILP} is much higher than that of the $DetermineRobustness$ algorithm (e.g. for $k$-out random digraphs with $k=4$). It is not immediately clear why this is the case; future work will investigate graph characteristics which affect the convergence time of the MILP formulations. Finally, for small values of $n$ (e.g. $n \in \{7,8\}$) the average time for $DetermineRobustness$ is lower than the average time for the \emph{$(r,s)$-Rob. MILP}. This likely reflects that it may be quicker to simply test all unique nonempty, disjoint subsets in these cases (966 for $n = 7$, 3025 for $n = 8$) than to incur computational overhead associated with solving the MILP formulations.  We point out that, with a few exceptions, the difference in this case is small: the average convergence time for both algorithms is generally under $10^{-1}$ seconds for $n \in \{7,8\}$. 


 The second simulation set compares the performance of four algorithms which determine only the value of $\rmax(\D)$ for digraphs. These include Algorithm \ref{alg:modified}, a modified version of $DetermineRobustness$ which determines $\rmax(\D)$, the MILP formulation from Theorem \ref{thm:rrobust} (denoted \emph{r-Rob. MILP}), the lower bound MILP formulation from Theorem \ref{thm:lowerbound} (denoted \emph{r-Rob. Lower Bnd}), and the upper bound MILP formulation from Theorem \ref{thm:twocase} (denoted \emph{r-Rob. Upper Bnd}). These algorithms are tested on the four types of graphs described above with values of $n$ ranging from $7$ to $15$. Additionally, the MILP formulations are tested on digraphs with values of $n$ ranging from $17$ to $25$. Again, 100 graphs per graph type and combination of $n$ and $p$ (or $n$ and $k$, depending on the respective graph type) are randomly generated, and the algorithms are run on each graph. Overall, 10,800 graphs are analyzed with Algorithm \ref{alg:modified} and 16,800 graphs are analyzed by each of the three MILP formulations. The time for each algorithm to determine $\rmax(\D)$ is averaged for each combination of $n$, $p$ or $k$, and graph type. The average, minimum, and maximum times per combination are plotted in Figure \ref{fig:r_sims}.
A time limit of $10^3$ seconds is again imposed on all three of the MILP formulations, but out of the 16,800 graphs tested there are no instances where this time limit was violated.

Some of the same patterns as in the first set of simulations (with the $DetermineRobustness$ and \emph{$(r,s)$-Rob. MILP} algorithms) are evident in the second set of simulations. The \emph{Mod. Det. Rob.} algorithm also terminates if a pair of subsets $S_1$ and $S_2$ are found which are both 0-reachable, which is likely the reason for the small minimum computation time of this algorithm for several of the graphs. \emph{Mod. Det. Rob.} generally has a lower average computational time for $n \in \{7,8\}$, again likely due to the speed of checking the relatively low number of unique nonempty, disjoint subset pairs as compared to solving the MILPs. It is not clear why the \emph{$r$-Rob. Upper Bnd} MILP exhibits high average and maximum computational times for the $k$-out random digraphs. Future work will further analyze graph characteristics which negatively affect the convergence time of the MILP formulations.



\begin{figure*}[h]
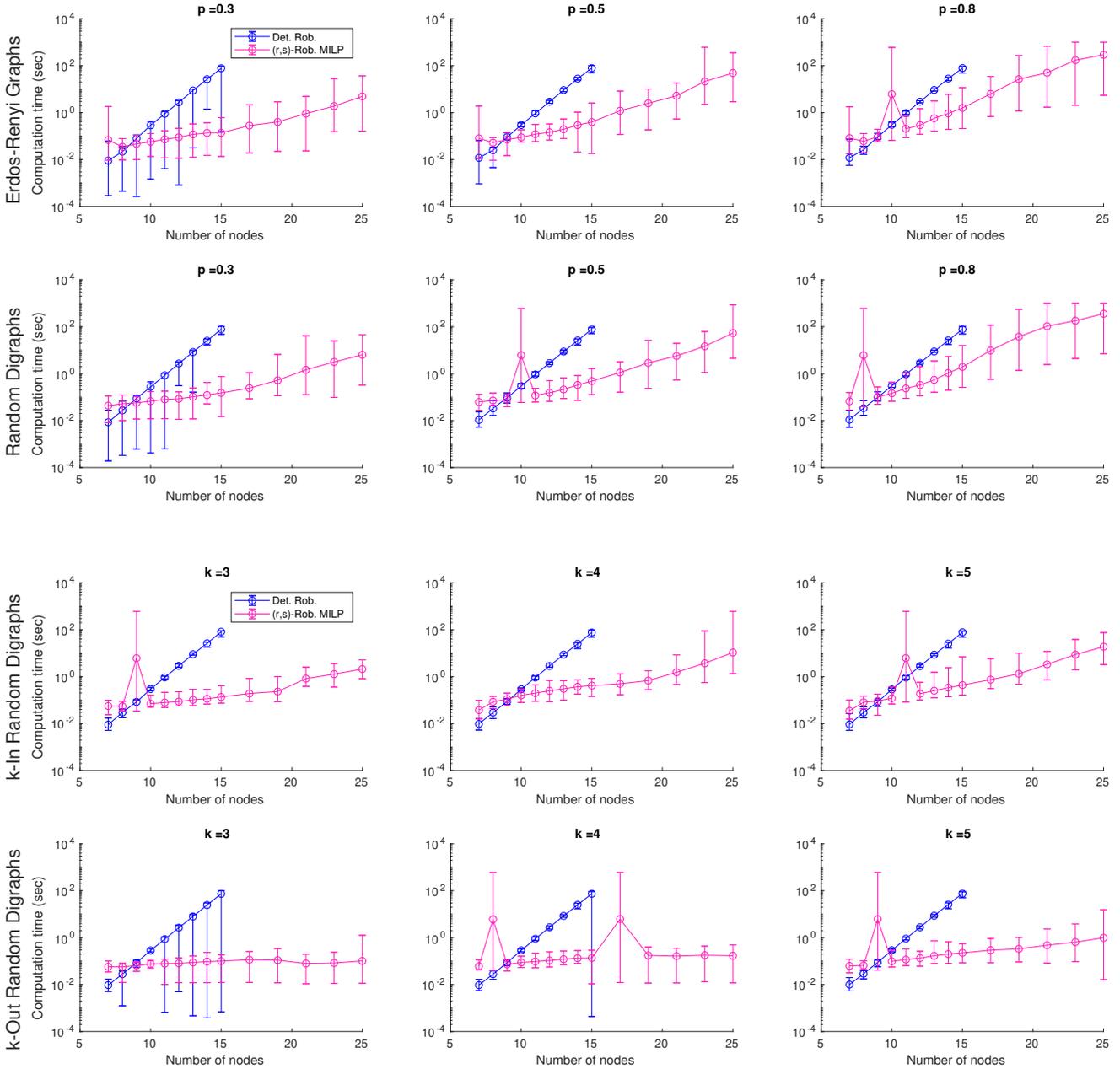
%

\begin{subfigure}%

\hspace*{-1.45cm}\includegraphics[width=\simsize\textwidth]{randomgraphs_r_s.eps}
\end{subfigure}%
\begin{subfigure}%

\hspace*{-1.45cm}\includegraphics[width=\simsize\textwidth]{kgraphs_r_s.eps}
\end{subfigure}%
\caption{
{\scriptsize
Comparison of $DetermineRobustness$ to \emph{$(r,s)$-Rob. MILP} (Algorithm \ref{alg:MILP}). The interpolating lines and circles represents the average computation time in seconds over 100 digraphs for each value of $n$, the upper and lower lines represent the maximum and minimum computation times, respectively, over the 100 trials for each $n$. Note that \emph{$(r,s)$-Rob. MILP} actually solves \emph{two} MILPs sequentially: one to find $\rmax(\D)$, and one to find $\smax(\rmax(\D))$.
}
}%
\label{fig:r_s_sims}%
\end{figure*}%
\begin{figure*}[h]
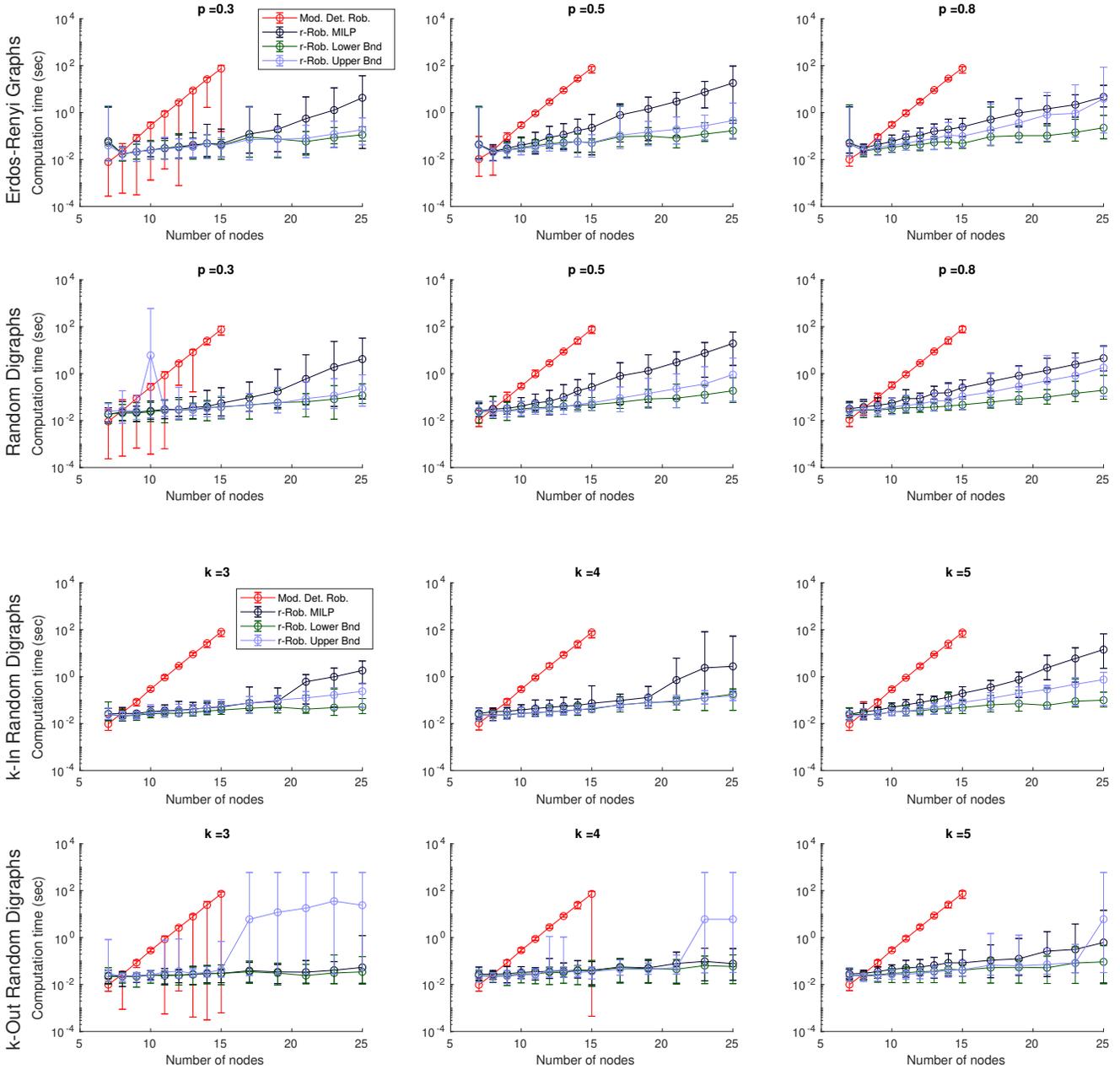
%

\begin{subfigure}%

\hspace*{-1.45cm}\includegraphics[width=\simsize\textwidth]{randomgraphs_r.eps}
\end{subfigure}%
\begin{subfigure}%
\centering
\hspace*{-1.45cm}\includegraphics[width=\simsize\textwidth]{kgraphs_r.eps}
\end{subfigure}%
\caption{
{\scriptsize
Comparison of the \emph{Mod. Det. Rob.} algorithm (Algorithm \ref{alg:modified}) which determines $\rmax(\D)$  to three MILP formulations. The first MILP formulation labeled \emph{r-Rob. MILP} is an implementation of Theorem \ref{thm:rrobust} and calculates $\rmax(\D)$ exactly. The MILP formulation labeled \emph{r-Rob. Lower Bnd} is an implementation of Theorem \ref{thm:lowerbound} and calculates a lower bound on $\rmax(\D)$. The MILP formulation labeled \emph{r-Rob. Upper Bnd} is an implementation of Theorem \ref{thm:twocase} and calculates an upper bound on $\rmax(\D)$. The interpolating lines and circles represents the average computation time over 100 digraphs for each value of $n$, the upper and lower lines represent the maximum and minimum computation times, respectively, over the 100 trials for each $n$.
}
}%
\label{fig:r_sims}%
\end{figure*}%

\vspace{-1em}

\section{Conclusion}
\label{sec:conclusion}

This paper presents a novel approach on determining the $r$- and $(r,s)$-robustness of digraphs using mixed integer linear programming (MILP). The advantages of the MILP formulations and branch-and-bound algorithms over prior algorithms are discussed, and the performance of the MILP methods to the $DetermineRobustness$ algorithm is compared.

Much work remains to be done in the area of robustness determination. The results in this paper merely open the door for the extensive literature on mixed integer programming to be applied to the robustness determination problem. Future work will focus on applying more advanced integer programming techniques to the formulations in this paper to yield faster solution times. In particular, the Laplacian matrix exhibits a high degree of structure and plays a central role in the MILP formulations presented in this paper. Future efforts will explore ways to exploit this structure to determine the robustness of digraphs more efficiently.


\bibliographystyle{plain}        
\bibliography{ACC2019}           



\appendix

\vspace{-1.5em}

\section{Description of Algorithm implementations}
\label{subsec:AppendixAlg}

\vspace{-.5em}

\begin{figure}
\label{fig:counterexample}
\centering
\begin{subfigure}
\centering
\includegraphics[width=.4\columnwidth]{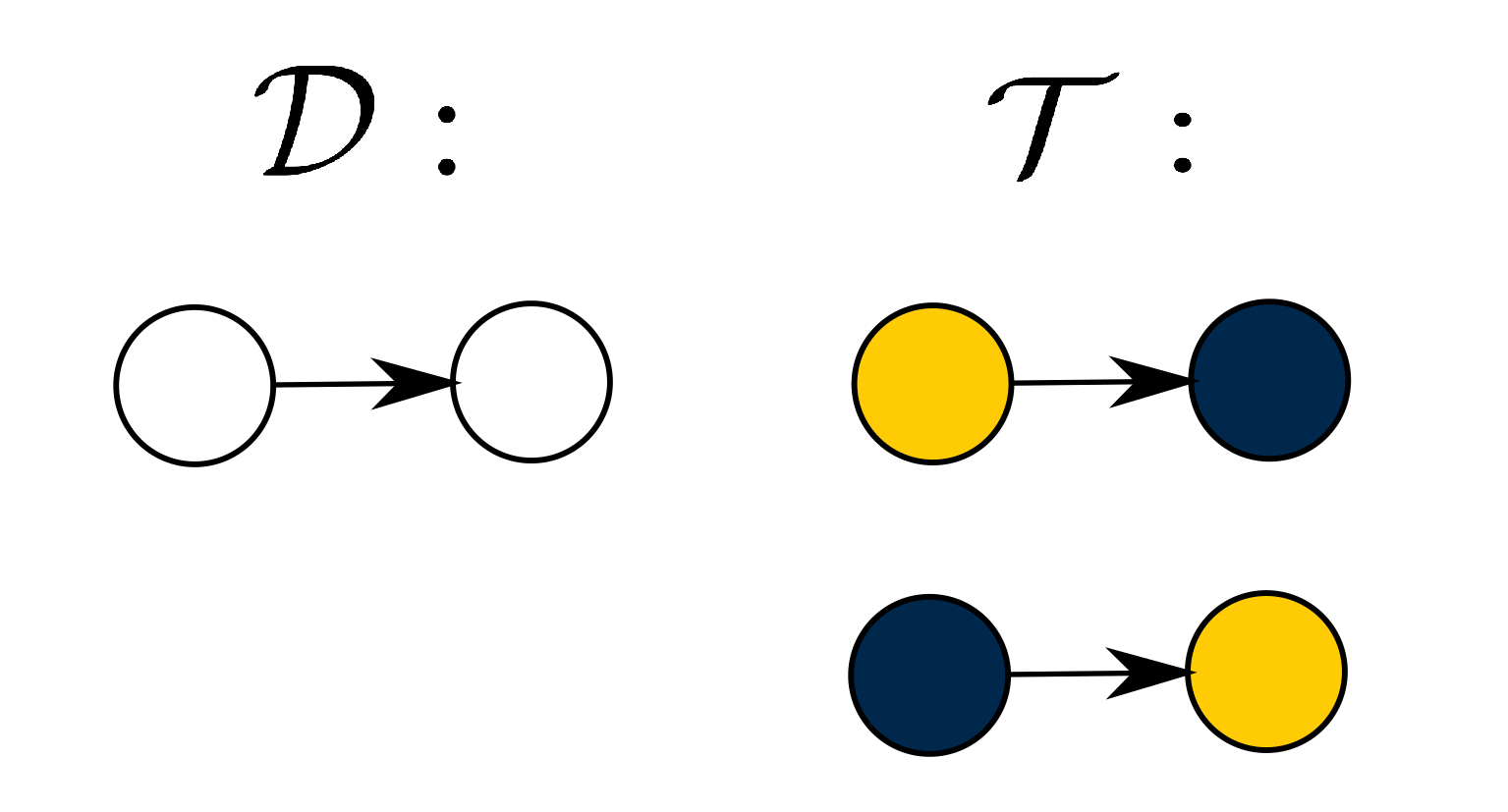}
\end{subfigure}
\begin{subfigure}
\centering
\includegraphics[width=.3\columnwidth]{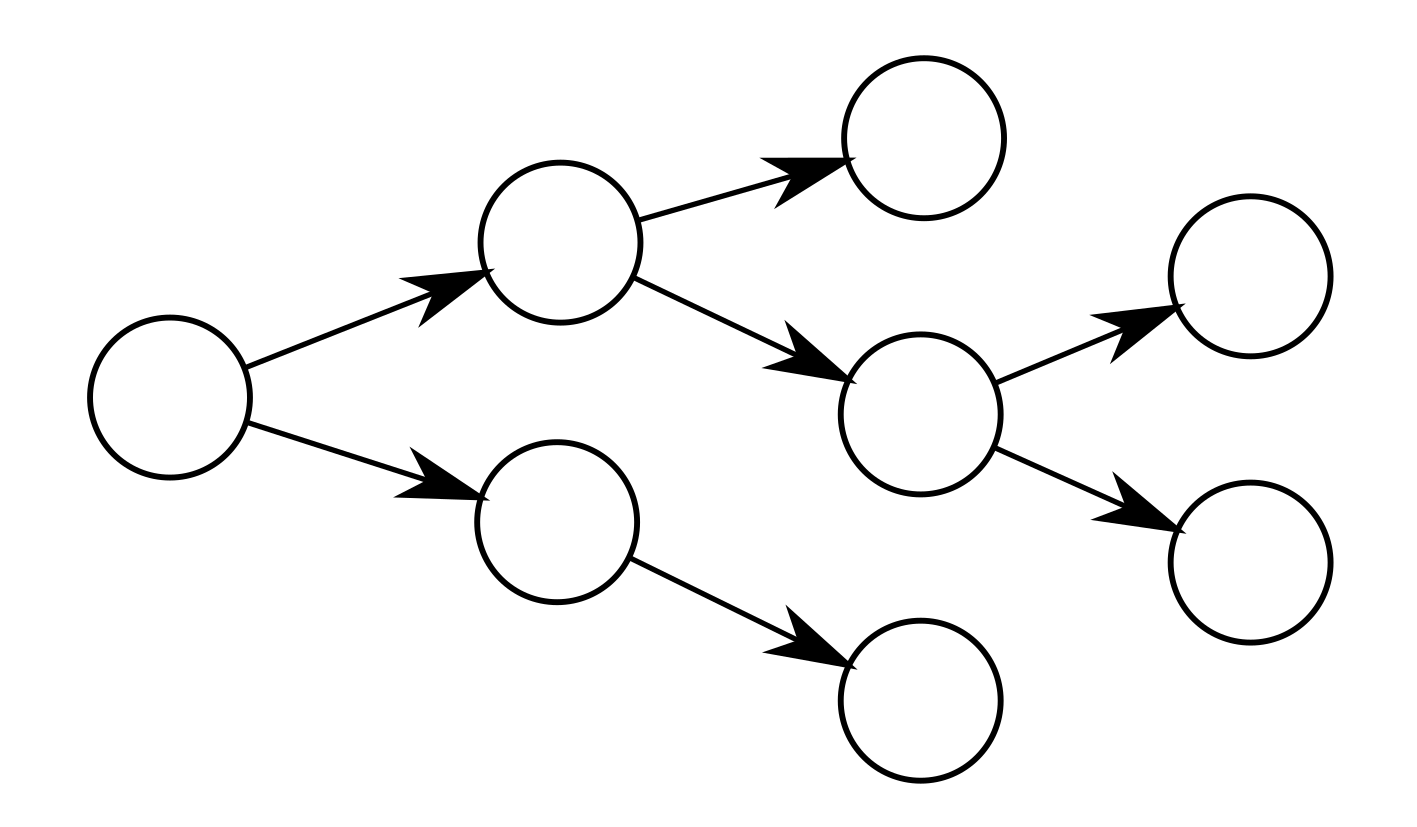}
\caption{
{\scriptsize
(Left) Example of a digraph which has $\delin(\D) = 0$ but which is 1-robust. The graph is depicted on the far left, and all possible $(S_1,S_2)$ pairs in $\Tc$ are depicted on the close left. (Right) Fig A.2. A rooted out-branching, where the in-degree of the root node (far left) is zero. All digraphs containing a rooted outbranching are at least $(1,1)$-robust \cite{leblanc2013resilient}.
}
}
\end{subfigure}
\end{figure}

This section gives additional details about the implementations of the algorithms tested in the Simulation.
Algorithm \ref{alg:DetRob} provides the details about the implementation of the $DetermineRobustness$ algorithm implemented in the simulations.
\begin{algorithm}
\caption{\cite{leblanc2013algorithms} \small{D\textsc{etermineRobustness}}$(A(\D))$ }\label{alg:DetRob}
\begin{algorithmic}[1]
\State \textbf{comment:} $A(\D)$ is the adjacency matrix of the graph.
\State $r \leftarrow \min\pth{\max\pth{\delta^{\text{in}}(\D),1},\ceil{\frac{n}{2}}}$
\State $s \leftarrow n$ 
\State \textbf{comment:} $\delta^{\text{in}}(\D)$ is the min. in-degree of nodes in $\D$
\ForAll{$k \larr 2$ to $n$}

    \ForAll {$K_i \in \mathcal{K}_k\ (i = 1,2,\ldots,\pmxs{n \\ k})$}
    \State \hspace{-1.9em} \textbf{comment:} $\mathcal{K}_k$ is the set of $\pmxs{n \\ k}$ unique subsets of $\V$
    \ForAll {$P_j \in \mathcal{P}_{K_i}\ (j = 1,2,\ldots,2^{k-1}-1)$}
    \State \hspace{-2.2em}  \begin{tabular}{l l} \textbf{comment:} &$\mathcal{P}_{K_i}$ is set of partitions of $K_i$ with \\
    & exactly two
    nonempty parts \end{tabular}
    \State \textbf{comment:} $\mathcal{P}_j = \{S_1,S_2 \}$
    \State ${isRSRobust \larr \text{R\textsc{obustholds}}}(A(\D),S_1,S_2,r,s)$
    \If{($isRSRobust == $ \textbf{false}) \textbf{and} $s > 0$}
        \State $s \larr s -1$
    \EndIf
    \While{$isRRobust == $ \textbf{false} \textbf{and} $(r > 0)$}
        \While{$isRSRobust == $ \textbf{false} \textbf{and} $(s > 0)$}
            \State $isRRobust$
            \State \hspace{2em}$\larr \text{R\textsc{obustholds}}(A(\D),S_1,S_2,r,s)$
            \If{\textbf{not} $isRSRobust$}
                \State $s \larr s-1$
            \EndIf
        \EndWhile

        \If{$isRSRobust == $ \textbf{false}}
            \State $r \larr r-1$
            \State $s \larr n$  
        \EndIf
    \EndWhile

    \If{$r == 0$}
        \State \textbf{return} $(r,s)$ \Comment{\emph{Implies $\rmax(\D) = 0$}}
    \EndIf
    \EndFor
    \EndFor
\EndFor
\State \textbf{return} $(r,s)$ \Comment{\emph{Returned values are $(\rmax(\D),\smax(\rmax(\D)))$}}
\end{algorithmic}
\end{algorithm}
One modification was made to the $DetermineRobustness$ algorithm to ensure accuracy of results. In the first line of the original $DetermineRobustness$ algorithm, $r$ is initialized with $\min \pth{\delta^{\text{in}}(\D), \ceil{n/2}}$. This yields incorrect results however for directed spanning trees where the in-degree of the root node is zero. Consider the digraph depicted in Figure \ref{fig:counterexample}. Here, $\delta^{\text{in}}(\D) = 0$, since the left agent has no in-neighbors. This implies that the original $DetermineRobustness$ algorithm would initialize $r \larr 0$ and return $(0,n)$ as the values of $(\rmax(\D),\smax(\rmax(\D)))$. However, Figure \ref{fig:counterexample} shows that for all nonempty, disjoint subsets $S_1$ and $S_2$, at least one is $1$-reachable. The depicted graph is therefore $(1,1)$-robust with $\rmax(\D) = 1$ and $\smax(\rmax(\D)) = 1$. 
In fact, initializing $r \larr \delta^{\text{in}}(\D)$ will always yield this error for any directed spanning tree where the in-degree of the root node is 0. This happens because any digraph is 1-robust if and only if it contains a rooted out-branching \cite[Lemma 7]{leblanc2013resilient}, yet $DetermineRobustness$ initializes $r \larr \delta^{\text{in}}(\D) = 0$ which results in termination at line 23.
In Algorithm \ref{alg:DetRob}, $r$ is instead initialized with $\min\pth{\max\pth{\delta^{\text{in}}(\D),1},\ceil{\frac{n}{2}}}$. This initializes $r \larr 1$ if $\delta^{\text{in}}(\D) = 0$, since it is still possible for the digraph to be $1$-robust.

To compare the MILP methodologies of this paper with the $DetermineRobustness$ algorithm, Algorithm \ref{alg:MILP} was used which determines the values of $(\rmax(\D), \smax(\rmax(\D)))$.
\begin{algorithm}
\caption{\small{$(r,s)$-\textsc{Rob. MILP}}}
\label{alg:MILP}
\begin{algorithmic}[1]
\State $r \larr \rmax(\D)$ from MILP in Theorem \ref{thm:rrobust}
\If {r == 0}
	\State $s \larr n$
\Else
	\If {$\delta^{\text{in}}(\D) \geq \floor{n/2} + r - 1$} \Comment{\emph{See Property 5.23 in \cite{leblanc2012thesis}}}
		\State $s \larr n$
	\Else
		\State $\smin(r) \larr $ from MILP in Theorem \ref{thm:rsgnl}
		\State $s \larr \smin(r) - 1$
	\EndIf
\EndIf
\State \textbf{return} $(r,s)$
\end{algorithmic}
\end{algorithm}
The first part of Algorithm \ref{alg:MILP} uses the formulation in Theorem \ref{thm:rrobust} to determine $\rmax(\D)$. If $\rmax(\D) = 0$, then $s \larr n$ and the algorithm returns $(r,s)$. If $\rmax(\D) > 0$, then the algorithm determines in line 5 whether the minimum in-degree of the graph $\delin(\D) \geq \floor{n/2} + r - 1$. By Property 5.23 in \cite{leblanc2012thesis}, if this is satisfied then the digraph is $(r,s)$-robust for all $1 \leq s \leq n$. Since for $r > 0$ the MILP in Theorem \ref{thm:rsgnl} is infeasible if and only if $s = n$, this test attempts to help the MILP solver avoid a fruitless search for a feasible solution if $s$ is indeed equal to $n$. This test works for graphs with large minimum in-degrees (e.g. complete graphs), but since it is a sufficient condition only it may not always detect when $\smax(r) = n$. Determining a more rigorous test to determine when $\smax(r) = n$ is left for future work.
Finally, if the test in line 5 fails then the MILP formulation in Theorem \ref{thm:rsgnl} is performed to determine $\smin$. Since $\smax(r) = \smin(r) - 1$, the value of $\smax(r)$ is stored in $s$ and the algorithm returns $(r,s)$.

The MILP algorithms in Section \ref{sec:rrobust} consider $r$-robustness, which is equivalent to $(r,1)$-robustness \cite[Property 5.21]{leblanc2013algorithms}, \cite[Section VII-B]{leblanc2013resilient}. Since they effectively do not consider values of $s$ greater than $1$, it is unfair to compare them directly with the $DetermineRobustness$ algorithm. Algorithm \ref{alg:modified} is a modified version of $DetermineRobustness$ which only considers $(r,1)$-robustness. This is accomplished by initializing $s \larr 1$ in lines 2 and 20 instead of $s \larr n$. Algorithm \ref{alg:modified} is labeled \emph{Mod. Det. Rob.} in the simulation legends.

\begin{algorithm}
\caption{\small{Modified version of D\textsc{etermineRobustness}}}\label{alg:modified}
\begin{algorithmic}[1]
\State \textbf{comment:} $A(\D)$ is the adjacency matrix of the graph
\State $r \leftarrow \min\pth{\max\pth{\delta^{\text{in}}(\D),1},\ceil{\frac{n}{2}}}$
\State $s \leftarrow 1$ \Comment{\emph{(Different than Alg. 3.2 in \cite{leblanc2013algorithms})}}
\State \textbf{comment:} $\delta^{\text{in}}(\D)$ is the min. in-degree of nodes in $\D$
\ForAll{$k \larr 2$ to $n$}
    \State \hspace{-1.9em} \textbf{comment:} $\mathcal{K}_k$ is the set of $\pmxs{n \\ k}$ unique subsets of $\V$
    \ForAll {$K_i \in \mathcal{K}_k\ (i = 1,2,\ldots,\pmxs{n \\ k})$}
    \ForAll {$P_j \in \mathcal{P}_{K_i}\ (j = 1,2,\ldots,2^{k-1}-1)$}
    \State \hspace{-2.2em}  \begin{tabular}{l l} \textbf{comment:} &$\mathcal{P}_{K_i}$ is set of partitions of $K_i$ into $S_1$ \\
    &and $S_2$ \end{tabular}
    \State ${isRSRobust \larr \text{R\textsc{obustholds}}}(A(\D),S_1,S_2,r,s)$
    \If{($isRSRobust == $ \textbf{false}) \textbf{and} $s > 0$}
        \State $s \larr s -1$
    \EndIf
    \While{$isRRobust == $ \textbf{false} \textbf{and} $(r > 0)$}
        \While{$isRSRobust == $ \textbf{false} \textbf{and} $(s > 0)$}
            \State $isRRobust$
            \State \hspace{2em}$\larr \text{R\textsc{obustholds}}(A,S_1,S_2,r,s)$
            \If{\textbf{not} $isRSRobust$}
                \State $s \larr s-1$
            \EndIf
        \EndWhile

        \If{$isRSRobust == $ \textbf{false}}
            \State $r \larr r-1$
            \State $s \larr 1$  \Comment{\emph{(Diff. than Alg. 3.2 in \cite{leblanc2013algorithms})}}
        \EndIf
    \EndWhile

    \If{$r == 0$}
        \State \textbf{return} $r$
    \EndIf
    \EndFor
    \EndFor
\EndFor
\State \textbf{return} $r$
\end{algorithmic}
\end{algorithm}

\end{document}